\documentclass[12pt,preprint]{aastex}

\newcommand{\arcm}{\ifmmode {' }\else $' $\fi}
\newcommand{\arcs}{\ifmmode {'' }\else $'' $\fi}

\newcommand{\lapp}{$_<\atop{^\sim}$}
\newcommand{\gapp}{$_>\atop{^\sim}$} 
\slugcomment{}

\shortauthors{Rhode et al.} \shorttitle{Globular Cluster Systems of
  NGC~891 and NGC~4013}

\begin{document}

\title{WIYN Imaging of the Globular Cluster Systems \\of the Spiral Galaxies NGC~891 and NGC~4013}

\author{Katherine L. Rhode, Jessica L. Windschitl, and Michael D. Young}
\affil{Department of Astronomy, Indiana University, Swain West 319, 727 East Third
  Street, Bloomington, IN 47405-7105, USA; rhode@astro.indiana.edu} 

\begin{abstract}
We present results from a WIYN 3.5-m telescope imaging study of the
globular cluster (GC) systems of the edge-on spiral galaxies NGC~891
and NGC~4013.  We used the 10$\arcm$x10$\arcm$ Minimosaic Imager to
observe the galaxies in $BVR$ filters to projected radii of
$\sim$20~kpc from the galaxy centers.  We combined the WIYN data with
archival and published data from the $WFPC2$ and Advanced Camera for
Surveys on the {\it Hubble Space Telescope} to assess the
contamination level of the WIYN GC candidate sample and to follow the
GC systems further in toward the galaxies' centers.  We constructed
radial distributions for the GC systems using both the WIYN and HST
data. The GC systems of NGC~891 and NGC~4013 extend to 9$\pm$3~kpc and
14$\pm$5~kpc, respectively, before falling off to undetectable levels
in our images.  We use the radial distributions to calculate global
values for the total number ($N_{GC}$) and specific frequencies ($S_N$
and $T$) of GCs. NGC~4013 has $N_{GC}$ $=$ 140$\pm$20, $S_N$ $=$
1.0$\pm$0.2 and $T$ $=$ 1.9$\pm$0.5; our $N_{GC}$ value is $\sim$40\%
smaller than a previous determination from the literature.  The $HST$
data were especially useful for NGC~891, because the GC system is
concentrated toward the plane of the galaxy and was only weakly
detected in our WIYN images. Although NGC~891 is thought to resemble
the Milky Way in terms of its overall properties, it has only half as
many GCs, with $N_{GC}$ $=$ 70$\pm$20, $S_N$ $=$ 0.3$\pm$0.1 and $T$
$=$ 0.6$\pm$0.3. We also calculate the galaxy-mass-normalized number
of blue (metal-poor) GCs in NGC~891 and NGC~4013 and find that they
fall along a general trend of increasing specific frequency of blue
GCs with increasing galaxy mass.  Given currently available resources,
the optimal method for studying the global properties of extragalactic
GC systems is to combine $HST$ data with wide-field, ground-based
imaging with good resolution.  The results here demonstrate the
advantage gained by using both methods when possible.
\end{abstract}


\section{Introduction}
\label{section:introduction}

Globular clusters (GCs) are among the most useful stellar populations
to study in galaxies.
%in the local
%universe.  
These compact star clusters each contain $\sim$10$^3$ $-$ 10$^6$
%thousands to millions
of stars that appear to have formed over a relatively short period of
time and under similar conditions (e.g., Ashman \& Zepf 1992; Ashman
\& Zepf 1998; West et al.\ 2004; Brodie \& Strader 2006; Griffen et
al.\ 2009; Gnedin 2010).  GCs are luminous (Milky Way GCs typically
have $M_V$ $\sim$ $-$10 to $-$5; Harris 1996), often many Gyr old
(e.g., Chaboyer et al.\ 1998, Brodie \& Strader 2006), appear to be
associated with major star formation episodes in galaxies (Brodie \&
Strader 2006, Bastian 2008), and may be forming in nearby galaxy
mergers and/or starbursts (e.g., Whitmore \& Schweizer 1995, de~Grijs
et al.\ 2001).  GCs are detected in virtually all giant galaxies that
have been imaged to sufficient depth to cover a significant fraction
of the GC luminosity function (GCLF; Harris 1991, Brodie \& Strader
2006).
%or just ``studied observationally.''
Together, these properties mean that GCs can be used as a stellar
record of the major events associated with the origin and evolution of
the galaxies that host them.  The numbers, spatial distributions,
luminosities,
%colors (which can give an indication of their
metallicities, and kinematics of the GCs
%can be used to help establish 
provide fundamental information about the star formation and assembly
histories of giant galaxies \citep{brodie06}.
%observational constraints on the star formation
%and assembly history of their parent galaxy \citep{brodie06}.

Motivated primarily by 
these ideas,
%the potential utility of GCs as records of galaxy formation and
%  evolution, 
we have been carrying out a wide-field CCD imaging survey of the GC
  systems of giant spiral, elliptical, and S0 galaxies in the local
  universe (distances to $\sim$20~Mpc).  The spiral galaxies we target
  are all edge-on or nearly edge-on, to make it easier to identify
  bona fide GCs and less likely that we will mistake knots of star
  formation or open clusters in the galaxy disk for GCs.  We use
  large-format and mosaic CCD detectors to image the target galaxies
%(which have distances of $\sim$7$-$20~Mpc)
in multiple broadband filters and select as GC candidates the point
sources around the galaxies with the expected magnitudes and colors.
%based on their magnitudes
%and colors.  
We use the GC candidate lists and galaxy images to estimate the global
properties of each galaxy's GC system --- namely, the total number and
specific frequency of GCs as well as the spatial and color
distribution of the overall system. Since GC systems of nearby
galaxies are extended (often covering tens of arc minutes from the
galaxy center), wide-field imaging is necessary to cover enough of the
GC system to allow for direct, accurate determinations of these
quantities. 

The general goal of the survey is to increase the number of galaxies
for which accurate measurements of the global GC system properties
exist, so that these measurements can in turn be used to test
scenarios for the formation and evolution of giant galaxies.  The
details of the design and methods of the survey, as well as the
results from the first ten galaxies studied, are given in Rhode \&
Zepf (2001, 2003, 2004; hereafter RZ01, RZ03, RZ04) and in Rhode et
al.\ (2005, 2007; hereafter R05, R07).  Another motivation for the
survey is to use the GC candidate lists as target lists for follow-up
spectroscopy, in order to measure the radial velocities of the GCs
around each galaxy.  The kinematics of the GCs provide additional
insight into the origin of each host galaxy and the radial velocities
can also be used to quantify the galaxy's mass distribution (e.g.,
Bergond et al. 2006, Bridges et al.\ 2007).

In this paper, we present results for two more galaxies imaged as part
of the wide-field survey: NGC~4013 and NGC~891, both edge-on Sb spiral
galaxies.  Some fundamental properties of the galaxies are summarized
in Table~\ref{table:properties}. To analyze these galaxies' GC
systems, we combined wide-field mosaic CCD data obtained with the WIYN
3.5-m telescope\footnote{The WIYN Observatory is a joint facility of
the University of Wisconsin, Indiana University, Yale University, and
the National Optical Astronomy Observatory.}  with previously
published results from {\it Hubble Space Telescope (HST)} imaging
studies of the inner parts of the galaxies' GC systems. Ground-based
imaging and HST imaging each have advantages and disadvantages as
methods for studying extragalactic GC systems.  Ground-based imaging
with large-format CCDs often enables one to image all, or nearly all,
of the galaxy's GC system in a single pointing.  However, limited
resolution makes it difficult to distinguish GCs from distant
background galaxies and to detect GCs close in to the center of the
parent galaxy, especially in the case of a spiral galaxy with a dusty
disk and/or active regions of star formation. HST's superb resolution
allows one to detect GCs further in toward the centers of their host
galaxies, to distinguish GCs from compact background galaxies, and in
some cases to study the sizes and structural parameters of the
individual GCs (e.g., Spitler et al.\ 2006, Forbes et al.\ 2010,
Harris et al.\ 2010).  On the other hand, the limited fields-of-view
of the HST imagers means that only a fraction of the GC system is
typically observed (see discussions in Ashman \& Zepf 1998 and Brodie
\& Strader 2006), and properties such as the overall spatial distribution
and specific frequency of the GC system can be undetermined or highly
uncertain without supplemental data or assumptions (e.g., Goudfrooij
et al.\ 2003, Chandar, Whitmore, \& Lee 2004, Peng et al.\ 2008,
Cantiello, Brocato, \& Blakeslee 2009).

The results presented here for NGC~4013 and NGC~891
%of the current study 
demonstrate the efficacy of combining wide-field ground-based data
with data from HST to study GC systems of giant galaxies in the local
universe.  The following section describes the WIYN observations and
initial image reduction steps.  Section~\ref{section:detection}
presents the techniques used to detect and select GC candidates in the
WIYN images.  Section~\ref{section:analysis2} describes additional
analysis steps we performed to assess the completeness and
contamination level of the WIYN data and ends with a description of
how we combined the published HST results for these galaxies with the
WIYN data. The overall results for the galaxies' GC systems are
described in Section~\ref{section:results}.  The paper ends with a
summary of our main conclusions in Section~\ref{section:summary}.

\section{Observations and Image Reductions}
\label{section:reductions}

\begin{figure}
%\plotone{ngc891_bw.eps}
\plotone{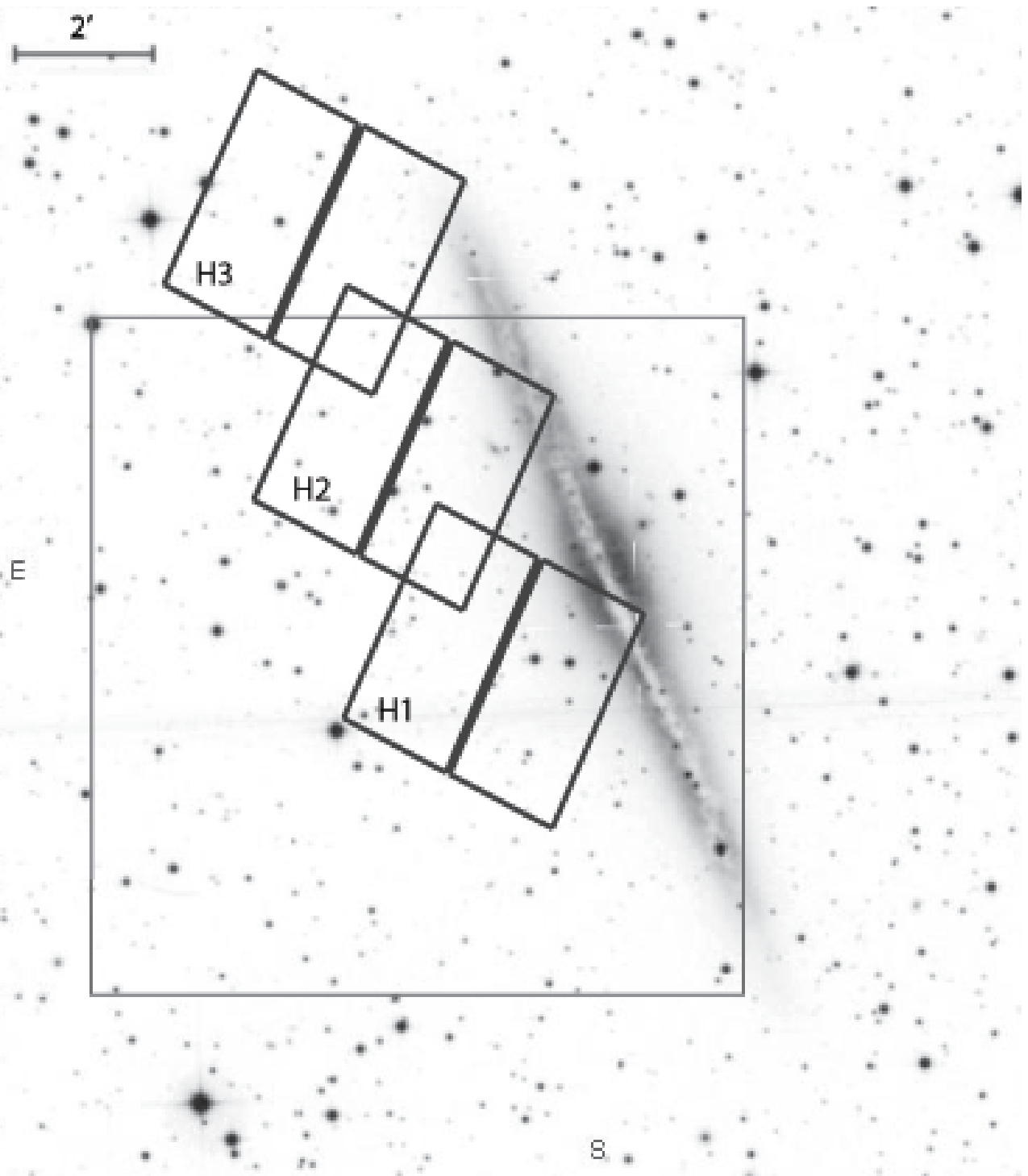}
\caption{\normalsize Digitized Sky Survey image of NGC~891 with the
  location of the WIYN pointing (large box) superimposed.  The three
  smaller boxes marked H1, H2, and H3 show the locations of the HST
  ACS fields analyzed by Harris et al.\ (2009) and discussed in
  Section~\ref{section:hst2}.} 
\label{fig:n891 pointing}
\end{figure}

\begin{figure}
%\plotone{ngc891_bw.eps}
\plotone{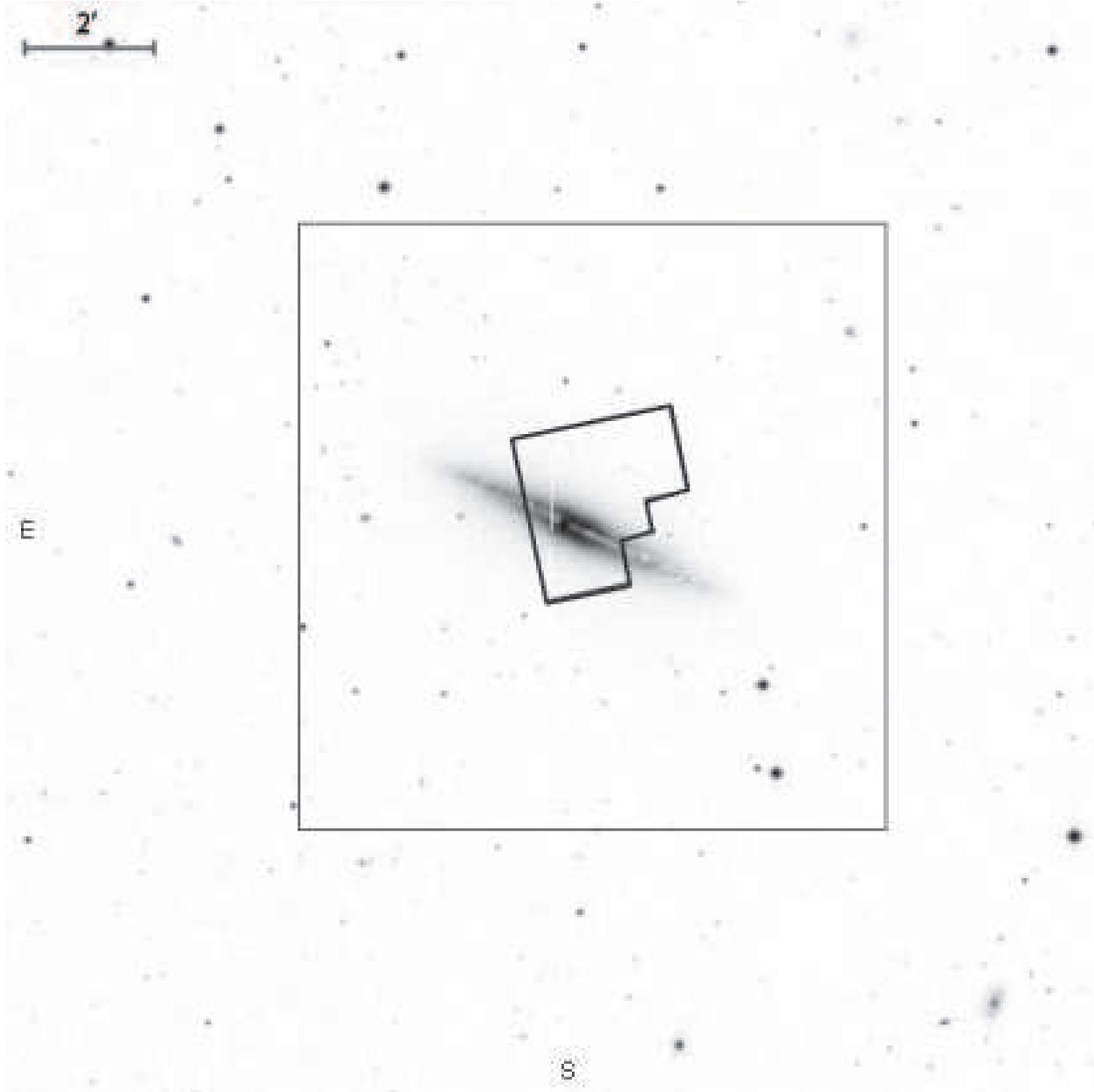}
\caption{\normalsize Digitized Sky Survey image of NGC~4013 with the
  location of the WIYN pointing (large box) superimposed. The HST
  WFPC2 pointing analyzed by Goudfrooij et al.\ (2003) and discussed
  in Section~\ref{section:hst2} is also marked.}
\label{fig:n4013 pointing}
\end{figure}

NGC~891 and NGC~4013 were imaged in three broadband filters ($BVR$)
with the Minimosaic camera on the 3.5-m WIYN telescope at Kitt Peak
National Observatory.  The WIYN Minimosaic is constructed of two
2048x4096-pixel CCDs with 0.14'' pixels and the field-of-view is
9.6$\arcm$ on each side.  NGC~891 is relatively nearby (see
Table~\ref{table:properties}) and thus was positioned on the edge of
the detector to maximize the radial coverage of its GC system;
NGC~4013 is more distant and was placed in the center of the
field. Figures~\ref{fig:n891 pointing} and \ref{fig:n4013 pointing}
show the locations of the WIYN fields, superposed on gray-scale images
from the Digitized Sky Survey\footnote{The Digitized Sky Surveys were
  produced at the Space Telescope Science Institute under
  U.S. Government grant NAG W-2166.The images of these surveys are
  based on photographic data obtained using the Oschin Schmidt
  Telescope on Palomar Mountain and the UK Schmidt Telescope. The
  plates were processed into compressed digital form with the
  permission of these institutions.}.  The WIYN observations analyzed
for this paper were acquired in 2001 January and 2009 March;
Table~\ref{table:observations} lists the dates, number of images, and
exposure times used for each galaxy pointing.

The sky conditions were photometric during the second night of the
2001 January observing run. On that night, images of standard stars
from Landolt (1992) were taken, along with one or more exposures of
NGC~891 and NGC~4013, in all three filters.
%ach of the $B$, $V$, and $R$
%filters and 
A set of photometric calibration coefficients (color terms and zero
point constants) was derived using the Landolt star observations taken
that night.  The color coefficient in the $V$ magnitude equation was
0.02$\pm$0.01. The color coefficients in the $B-V$ and $V-R$ color
equations were 1.02$\pm$0.01 and 1.05$\pm$0.03, respectively.  The
formal errors on the zero points in the $V$, $B-V$ and $V-R$
calibration equations ranged from 0.004 to 0.006, confirming that the
sky conditions that night were indeed photometric.

Images of the target galaxies taken on other, non-photometric nights
during the 2001 January run were calibrated by scaling them to a selected
image taken on the photometric night and then combining them with that
image (see below for details of the combining process). For NGC~4013,
a set of $V$ images was obtained in 2009 March; those images were
postcalibrated by calculating the zero-point offset in $V$ magnitude
between the final, combined 2009 March $V$ image and a single $V$ image
taken in 2001 January.

Initial reductions (overscan and bias level subtraction, flat-field
division) were performed on the Minimosaic images using standard
reduction tasks in the IRAF package MSCRED.  The tasks
\texttt{msczero}, \texttt{msccmatch}, and \texttt{mscimage} were then
used to transform the multi-extension FITS images into
single-extension images.  All of the images for a given galaxy were
aligned to match a single pointing.
%with a single reference image
%chosen from the set.  
A constant sky background level was computed and then subtracted from
each individual galaxy exposure.  Images of the same galaxy target,
taken in the same filter, were scaled to a common flux level and
combined using the IRAF task \texttt{imcombine} and the
\texttt{ccdclip} pixel rejection algorithm.  The constant sky
background level measured in the image used as a reference in the
scaling step was added to the final combined image. The end result was
one deep, stacked image per filter for each galaxy target. The mean
full-width at half-maximum of the point-spread function (FWHM PSF) in
the final combined images ranges from 1.0$\arcsec$$-$1.2$\arcsec$ for
NGC~891 and 0.7$\arcsec$$-$0.8$\arcsec$ for NGC~4013.
%
%n891_b_gcs.fits,FWHMPSF = 8.167 = 1.1''
%n891_r_gcs.fits,FWHMPSF = 7.134 = 1.0''
%n891_v_gcs.fits,FWHMPSF = 8.35 = 1.2''
%
%n4013_new_b_gcs.fits,FWHMPSF = 4.92 = 0.69''
%n4013_new_r_gcs.fits,FWHMPSF = 5.227 = 0.74''
%n4013_new_v_gcs.fits,FWHMPSF = 5.808 = 0.82''

\section{Detection and Analysis of the Globular Cluster System}
\label{section:detection}

\subsection{Source Detection}
\label{section:source detection}

GCs should appear in our ground-based WIYN Minimosaic images as a
slight overdensity of point sources surrounding the target galaxy.  To
detect and then analyze the GC systems of NGC~891 and NGC~4013, we
executed the same series of steps used in the previous studies that
are part of our wide-field survey (RZ01, RZ03, RZ04, R07).  We first
smoothed the combined images using a circular median filter with a
diameter equal to $\sim$7.25$-$7.5 times the mean FWHM PSF of the
image.  The smoothed images were subtracted from the original versions
to remove the diffuse light from the target galaxy and make it easier
to locate the point-source GCs.  The constant sky background level
present in the original images was added to the galaxy-subtracted
images; all subsequent analysis steps were performed on the
galaxy-subtracted images. The IRAF task \texttt{daofind} was used to
detect sources above a given signal-to-noise level in each frame.
%SNR between 4 and 5.5 times the sky noise
Areas of the images likely to produce spurious source detections --
i.e., near saturated stars and in the dust-obscured spiral disks of
the target galaxies -- were excluded from the detection process. A
final list of sources detected in all three filters was produced;
there were 774 objects in NGC~891 and 998 in NGC~4013.

\subsection{Removing Extended Sources}
\label{section:ext source cut}

Objects that appear extended in our $\sim$1$\arcsec$-resolution images
are likely to be background galaxies rather than GCs. To remove these
objects, we measured the FWHM of each source and plotted it versus the
source's instrumental magnitude.  Bright point sources form a tight
sequence around the mean FWHM value for each image, with increasing
scatter around the mean as the instrumental magnitude increases.  We
wrote software to select objects based on this pattern; the range of
acceptable FWHM values for point sources increases with source
magnitude. If a source failed the FWHM criteria for a point source of
a given magnitude in any one filter, it was removed from the list of
potential GC candidates. After this step, the number of sources
remaining was 490 for NGC~891 and 488 for NGC~4013.

\subsection{Aperture Photometry}
\label{section:photometry}

Aperture photometry was performed on the remaining point sources in
the $BVR$ images of NGC~891 and NGC~4013, using an aperture of radius
equal to the mean FWHM PSF for the specific image. An aperture
correction was derived for each image by measuring the mean difference
in the total magnitude and the magnitude within a one-FWHM-radius
aperture. The aperture corrections ranged from $-$0.229 to $-$0.308
with errors of 0.001 to 0.005.
%  
%N891 Aperture Corrections
%filter		aper corr	stddev		SEM
%B		-0.283933 	0.00531126 	0.00137136
%V		-0.308143 	0.014125 	0.00377507
%R		-0.285267 	0.00805753 	0.00208045
%
%N4013 Aperture Corrections
%filter		aper corr	stddev		SEM
%B	-0.270778 	0.00819722 	0.00273241
%V	-0.290923 	0.0167006 	0.00463191
%R       -0.228786       0.0104158       0.00278374
%
Final, calibrated total magnitudes were calculated for the point
sources in the images by applying the appropriate aperture correction
and photometric calibration coefficients, as well as a Galactic
extinction correction derived from the reddening maps of Schlegel,
Finkbeiner, \& Davis (1998). The Galactic extinction corrections for
NGC~891 were A$_B$$=$ 0.277, A$_V$$=$0.209, and A$_R$$=$0.168, and for
NGC~4013 were A$_B$$=$ 0.073, A$_V$$=$0.055, and A$_R$$=$0.045.

\subsection{Final Selection of GC Candidates}
\label{section:final selection}

The last step in the GC candidate selection process was to choose
objects with $BVR$ magnitudes and colors that match those expected for
GCs at the distance of the host galaxy.  We executed this step using
the same basic method described in the previous survey papers (RZ01,
RZ03, RZ04, and R07).  GCs were assumed to have
$M_V$$\sim$ $-$11 to $-$4, based on the GCLF of the Milky Way and
other giant galaxies \citep{az98}.  Objects within this magnitude
range (given the galaxy distances in Table~\ref{table:properties})
were checked to determine whether their $BVR$ colors were like those
expected for GCs with metallicities between [Fe/H] of
$-$2.5 and 0.0.  Specifically, sources with $B-V$ and $V-R$ values
(and corresponding photometric errors) that placed them within
2$-$3-$\sigma$ of the derived relation for Galactic GCs with the
specified metallicity range were accepted as GC candidates.

As explained in R07, the spiral galaxies included in our wide-field
survey typically have so few GC candidates (a few dozen, compared to
hundreds or thousands for the elliptical galaxies) that we carefully examine
each of the GC candidates as well as those that narrowly miss meeting
the above criteria.  We then adjust the selection criteria
accordingly, so as not to exclude objects that are likely to be GCs or
mistakenly include objects that are probably contaminating foreground
stars or background galaxies.  
%and adjust the selection criteria accordingly.  
We consider things like where objects lie in the $BVR$ color-color
plane, where they are located relative to the galaxy, and how the
objects appear in archival {\it HST} images (see
Section~\ref{section:hst1}).
%that might be available. 
%We then adjust the selection criteria slightly so as not to exclude
%objects that are likely to be GCs or mistakenly include objects that
%are probably contaminating foreground stars or background galaxies.
%We typically look at where objects lie in the $BVR$ color-color plane
%(since
%where they 
%we look at the subset of our GC candidates that appear in archival
%$HST$ images, and we look 

For NGC~4013, we executed the usual selection steps as described
above, accepting sources with $V$ magnitudes $\geq$19.9 and $BVR$
colors within 3-$\sigma$ of the $B-V$ vs.\ $V-R$ relation for Galactic
GCs.  We also decided to impose a faint magnitude limit of $V$ $<$
24.4, or $\sim$0.7-$\sigma$ past the GCLF peak, in order to minimize
the contamination from faint background galaxies that were not removed
in the extended source cut. The final set of GC candidates in NGC~4013
includes 83 objects. 

The GC candidate selection for NGC~891 was much more involved.  This
galaxy is at relatively low Galactic latitude ($-$17 degrees vs. the
more typical range for our targets, $+$40 to $+$70 degrees) and
therefore has a higher surface density of Galactic stars than typical
galaxy images from the survey.
%Galactic latitude of ngc891 = -17.4 degrees
% ngc2683 = 38.8
% ngc3556 = 56.3
% ngc4013 = 70.1
% ngc4157 = 65.5
% ngc7331 = -20.7
Imposing the usual luminosity threshold of $M_V$ $=$ $-$11 yielded a
significant population of bright sources in the GC candidate list
when, based on the shape of the Milky Way GCLF and that of other giant
galaxies, only a few real GCs would be expected at those magnitudes.
To account for this and to limit contamination from background
galaxies, we imposed more stringent $V$ magnitude criteria on both
ends of the luminosity function, choosing GC candidates in the range
20.7~$<$~$V$~$<$23.6 (extending from $\sim$1.3-$\sigma$ brighter than
the GCLF peak to 1-$\sigma$ fainter than the peak).  We also used a
2-$\sigma$ criterion for the $BVR$ color selection because analysis of
$HST$ images indicated that constraining the color cut would help
eliminate contaminants from the sample, while still allowing real GCs
to be selected.  Finally, we explicitly excluded 13 objects that
had count ratios in $HST$ images that indicated that they were
background galaxies or foreground stars. (The $HST$ data analysis is
discussed in Section~\ref{section:hst1}).

Lastly, we added a few sources to the final list for NGC~891. We
included four sources that were located very near the galaxy disk and
had colors that put them just outside the color selection box, in the
direction of the reddening vector.  We added two more sources
%to our sample that appeared to be
that had been excluded by the more stringent $V$ magnitude criteria,
but nevertheless appeared to be real GCs based on the {\it HST} study
we used to supplement the WIYN data (see Section~\ref{section:hst2}).
The final set of GC candidates for NGC~891 includes 43 objects.
%
%
%# Eliminate some sources that have ACS count ratios that indicate they
%# are foreground stars or background galaxies rather than GCs
%if ((id == 208)||(id == 215)||(id == 236)||(id == 246)){pass = 0}
%if ((id == 249)||(id == 254)||(id == 265)||(id == 281)){pass = 0}
%if ((id == 304)||(id == 311)){pass = 0}
%if ((id == 240)||(id == 306)||(id == 332)){pass = 0}

\begin{figure}
%\plotone{bvr_cc_plane_n4013_for_paper.ps}
\plotone{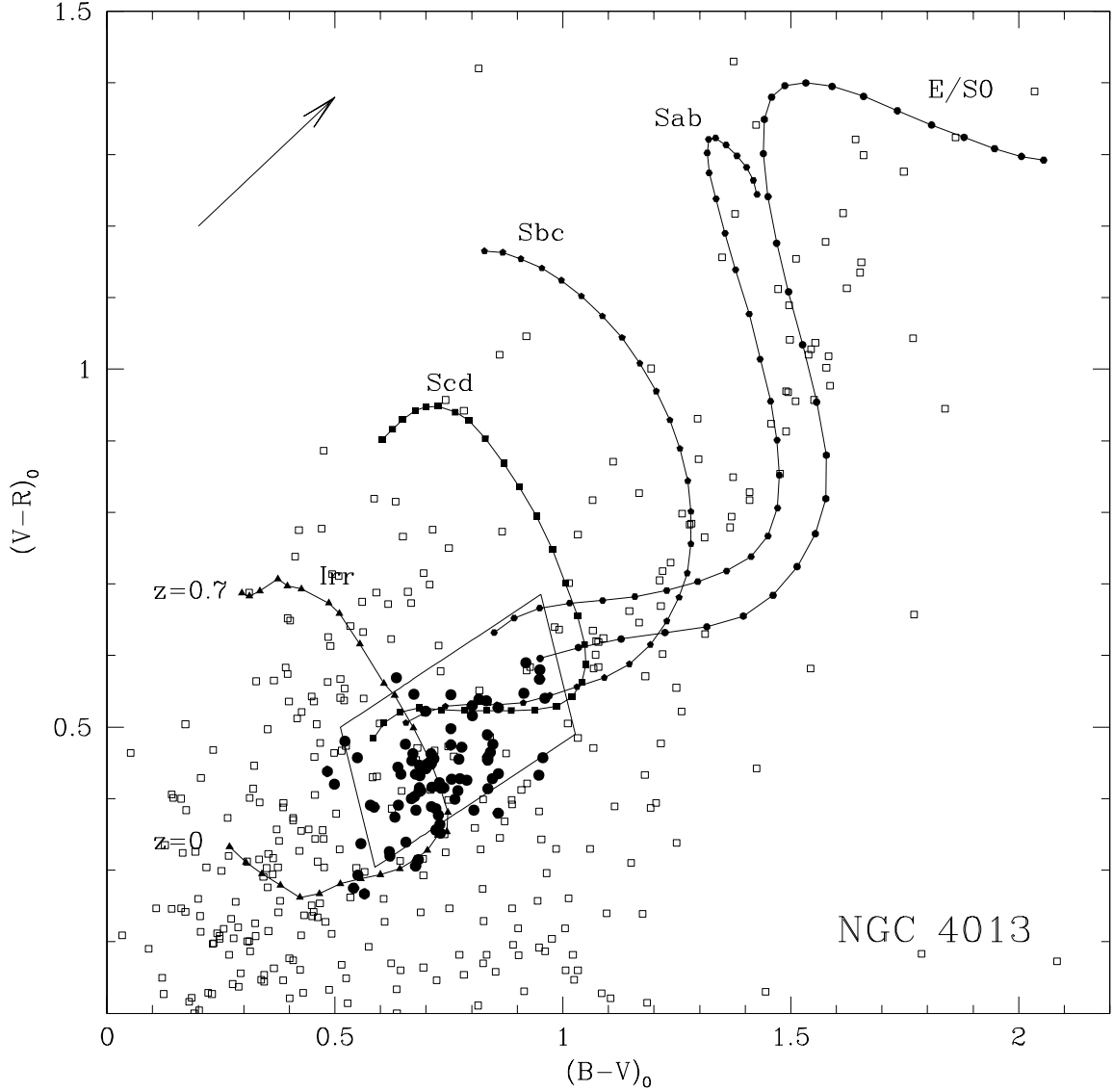}
\caption{\normalsize Color selection of GC candidates in NGC~4013.
  Open squares are the 488 point sources detected in all three
  filters; filled circles are the final set of 83 GC
  candidates. Because $V$ magnitude criteria were imposed, not all
  objects within the color selection box are GC candidates.  A
  reddening vector of length $A_V = 1$ mag appears in the upper
  left-hand corner. The box shows the boundaries of the color
  selection criteria.  The curves show where galaxies with different
  morphological types, located at redshifts from $z$ $=$ 0 to 0.7,
  would lie in this color-color plane.}
\label{fig:bvr n4013}
\end{figure}

\begin{figure}
%\plotone{bvr_cc_plane_n891_for_paper.ps}
\plotone{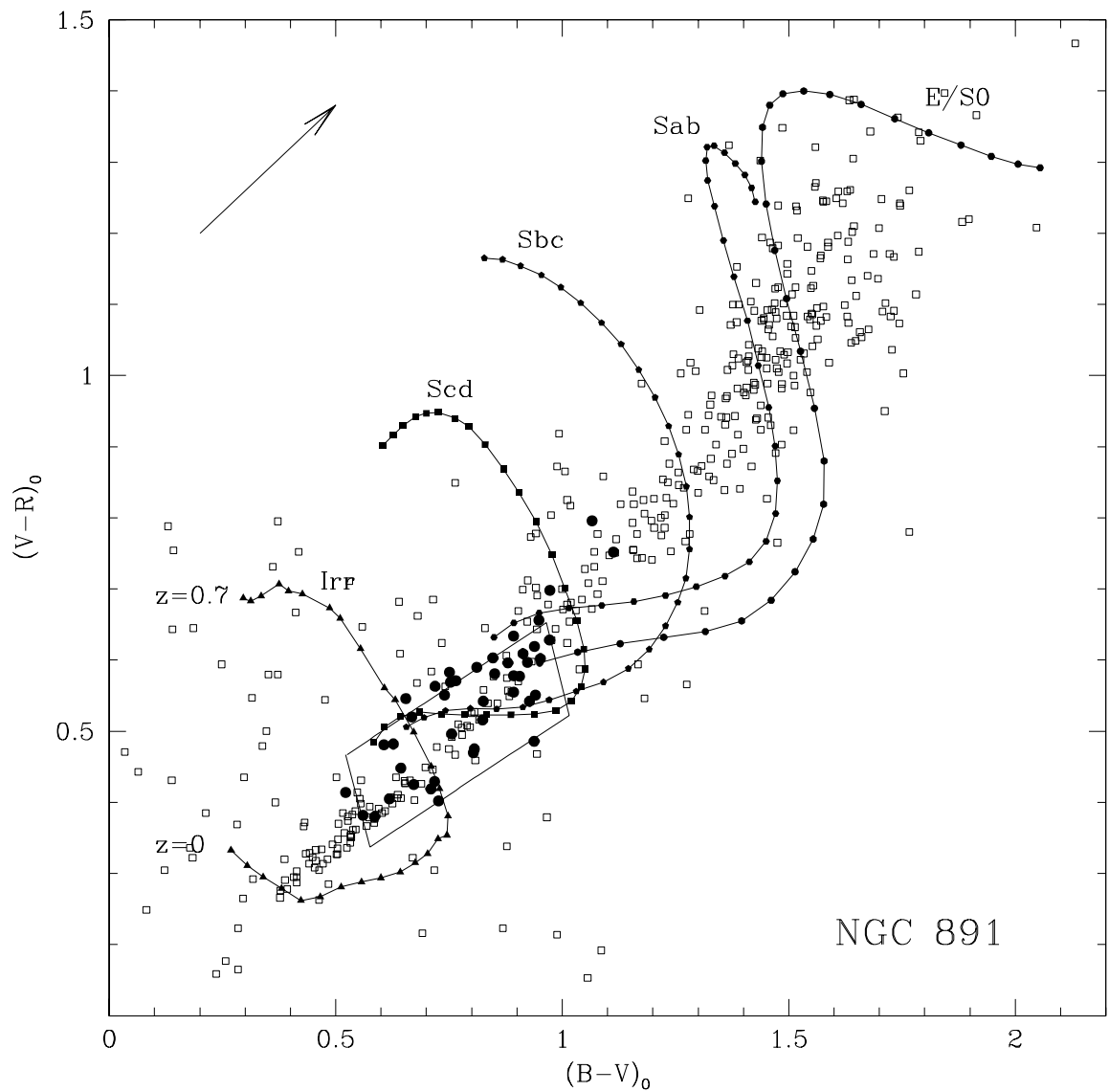}
\caption{\normalsize Color selection of GC candidates in NGC~891.
  Open squares are the 490 point sources detected in all three
  filters; filled circles are the final set of 43 GC
  candidates. Because $V$ magnitude criteria were imposed, not all
  objects within the color selection box are GC candidates. The other
  features of the plot (reddening vector, galaxy curves) are as in
  Figure~\ref{fig:bvr n4013}.}
% A reddening vector of length $A_V = 1$ mag appears in the upper
%  left-hand corner.  The box shows the boundaries of the color
%  selection criteria. The curves show where galaxies with different
%  morphological types, located at redshifts from $z$ $=$ 0 to 0.7,
%  would lie in this color-color plane.
\label{fig:bvr n891}
\end{figure}

Figures~\ref{fig:bvr n4013} and \ref{fig:bvr n891} show the results of
the color selection.  The point sources in each galaxy field that
appeared in all three filters are marked with open squares and the
final GC candidates are marked with filled circles.  The 2- or
3-$\sigma$-wide color selection boxes
%that were used 
are also marked.  Note that due to the $V$ magnitude criteria, not all
objects that appear within the color selection boxes are GC
candidates.
%To illustrate what types of sources
%would lie in the same region in the $BVR$ color-color plane as GCs,
For illustrative purposes, tracks are plotted in Figures~\ref{fig:bvr
  n4013} and \ref{fig:bvr n891} that show where galaxies of different
  morphological types (E/S0 through Irregular) at different redshifts
  would lie in the color-color plane. RZ01 details how the tracks were
  calculated. The distribution of sources looks markedly different in
  the NGC~4013 and NGC~891 fields and the galaxy tracks help explain
  why.  The NGC~4013 images are both deeper (see
  Section~\ref{section:completeness}) and located at higher Galactic
  latitude ($+$70 degrees) than the NGC~891 images.
% ngc891 = -17
% ngc4013 = 70.1
Consequently the figure is dominated by a large number of objects
located in the lower left part of the $BVR$ color-color plane for
NGC~4013; these are most likely faint, blue, low- to mid-$z$
background galaxies that are unresolved in our ground-based images.
By contrast the NGC~891 field has relatively few objects in the lower
left corner of the $BVR$ plane, and instead has a very well-populated,
narrow sequence of objects running diagonally through the plane; these
are likely Galactic foreground stars.

\begin{figure}
%\plotone{two_cmds_for_paper.ps}
\plotone{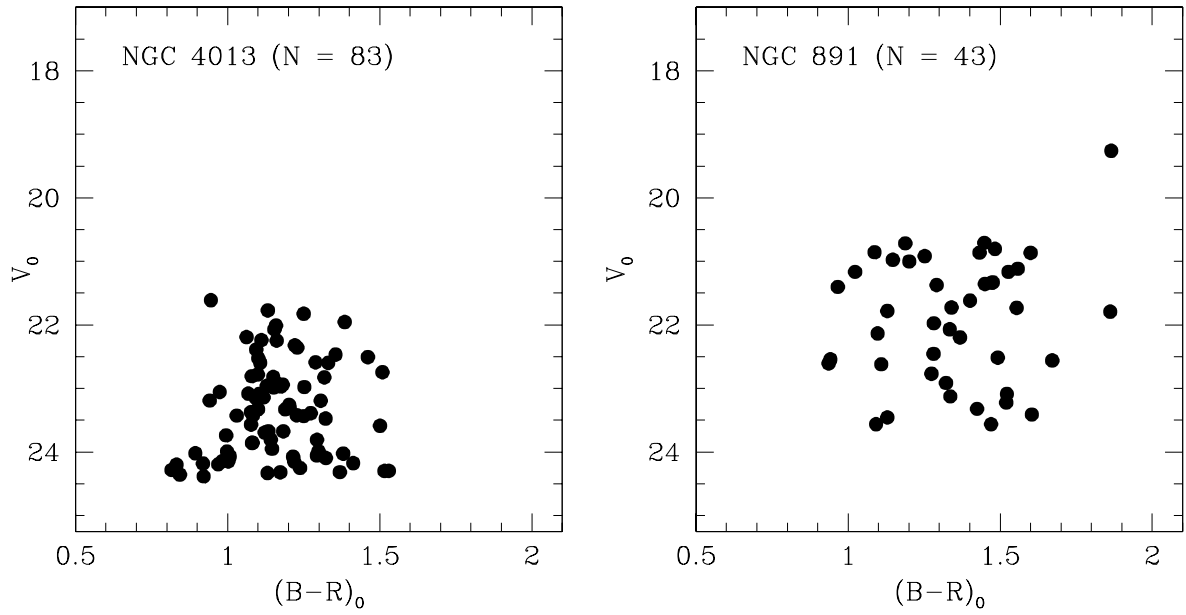}
\caption{\normalsize $V$ magnitude versus $B-R$ color for the final
  set of GC candidates in NGC~891 and NGC~4013. The magnitudes and
  colors have been corrected for Galactic extinction in the direction
  of the galaxy fields, but not for absorption internal to the
  galaxies themselves.}
\label{fig:cmds}
\end{figure}

The color-magnitude diagrams (CMDs) of the GC candidates for each
galaxy are shown in Figure~\ref{fig:cmds}.  To make use of information
from all three filters, the $V$ magnitudes vs.\ the $B-R$ colors of
the sources are plotted. The two very red sources with $B-R$~$\sim$1.9
in the NGC~891 CMD are very close to that galaxy's disk, which may
help explain their red colors; 
%are likely to be reddened due to their proximity to the galaxy's disk;
%very close to the disk of NGC~891 and likely to
the brighter of those two sources, with $V$~$\sim$~19, was added to
our sample based on its appearance as a legitimate GC in an $HST~ACS$
study from the literature (see Section~\ref{section:hst2}).

\section{Additional Analysis Steps}
\label{section:analysis2}

\subsection{Completeness Testing and Detection Limits}
\label{section:completeness}

We quantified the point-source detection limits of the WIYN Minimosaic
images by performing a series of completeness tests. In each test, 200
artificial sources, each with a point-spread-function that matched the
best-fit PSF for the image and a total magnitude within 0.2-magnitude
of a specified value, were added to a given Minimosaic image.  
%The sources had point-spread-functions (PSFs) that matched the intrinsic
%best-fit PSF for the image and total magnitudes within 0.2-mag of a
%specified value.  
The same detection steps used on the original images were then
executed and the fraction of artificial sources that was recovered 
%for that magnitude interval 
was recorded.  The process was repeated at 0.2-magnitude intervals
over a range of 5$-$6 magnitudes per filter.  This yielded a
measurement of completeness vs. magnitude for each filter and each
galaxy field.  The NGC~891 images are 50\% complete at $B=$25.2,
$V=$24.2, and $R=$24.1.  The images of NGC~4013 were taken under
better seeing conditions and accordingly are somewhat deeper, with
50\% detection limits at $B=$25.6, $V=$25.2, and $R=$24.8.

\subsection{Contamination}

A serious concern for ground-based imaging studies of extragalactic GC
systems is contamination from point-source stars and compact
galaxies masquerading as GCs in the candidate lists. We went through a
series of analysis steps to try to both quantify and correct for the
contamination from stars and galaxies in the GC candidate lists. 

\subsubsection{Model Prediction for Stellar Contamination}
\label{section:mendez model}

We used a Galactic structure model (Mendez \& van Altena 1996, Mendez
et al.\ 2000) to help us evaluate the level of stellar contamination
that exists in the GC candidate lists.  Quantities like the position
of the Sun within the Galaxy and the fraction of stars in the Galaxy's
structural components (disk, thick disk, and halo) are input to the
model; it then predicts the surface density of stars within a given
magnitude and color range and direction on the sky. Our experiments
with the model indicate that the predicted stellar surface density
does not strongly depend on the parameters like the Galaxy composition
and the solar distance and height above the Galactic disk.  The model
predicts that the surface density of Galactic stars with $V$
magnitudes and $B-V$ colors in the same range as the samples of GC
candidates is 0.06~arcmin$^{-2}$ for NGC~891 and 0.07~arcmin$^{-2}$
for NGC~4013.  Note that the value for the NGC~891 field is fairly
small, even though Figure~\ref{fig:bvr n891} suggests a large number
of stars in the field; this reflects the fact that we specifically
reduced
%strongly constrained / limited / actively specifically reduced 
the stellar contamination in the GC candidate sample for that galaxy
by limiting the sample to 20.7 $<$ $V$ $<$ 23.6. (If we instead impose
the usual $V$ magnitude threshold of $M_V$ $=$ $-$11, the $V$ range
becomes 18.6 $<$ $V$ $<$ 23.6 and the stellar contamination predicted
by the model is nearly four times larger, at 0.23~arcmin$^{-2}$.)

\subsubsection{Examining the WIYN GC Candidates in Archival HST Images}
\label{section:hst1}

Next we investigated the level of contamination in the GC lists due to
background galaxies. $HST$ can resolve many compact background
galaxies that appear as point sources in ground-based data, so we
analyzed archival $HST$ images of NGC~4013 and NGC~891 to determine
whether any of the WIYN GC candidates were actually galaxies.

For NGC~4013, we retrieved Wide Field Planetary Camera 2 (WFPC2)
images from the $HST$ data archive\footnote{Based on observations made
with the NASA/ESA {\it Hubble Space Telescope}, obtained from the data
archive at the Space Telescope Science Institute.  STScI is operated
by AURA, under NASA contract NAS 5-26555.}.  We analyzed two pointings
in F450W located $\sim$0.5$\arcm$ from the galaxy center ($HST$
program GO.8242, PI: Savage) and one pointing in F814W located
$\sim$0.7$\arcm$ from the galaxy center ($HST$ program GO.6685, PI:
Huizinga).  We requested that ``On-the-fly'' processing be applied to
the images before retrieval and we used the STSDAS task \texttt{
gcombine} to stack multiple pointings taken in the same
filter. Twenty-four of the WIYN GC candidates appear in one or more of
the WFPC2 frames.  Photometry was performed on these objects using
apertures 0.5- and 3-pixels in radius and the ratio of the flux within
the two apertures was computed.
%aperture of radius 0.5 and and another of 3 pixels and the ratio of
%the flux within the two apertures was computed.  
The ratio of counts in the larger aperture to smaller aperture should
be significantly larger for extended background galaxies than it is
for compact GC candidates (Kundu et al. 1999).  This test indicated
that none of the 24 WIYN GC candidates was a background galaxy.  The
24 candidates were inspected visually to confirm the results of the
count ratio test.  Note that this does {\it not} necessarily mean that
there is zero contamination from galaxies in the WIYN sample,
%because small-number statistics are a factor; we could only examine
%$\sim$30\% of the total WIYN sample.
just that the specific subset of candidates ($\sim$30\% of the total)
in the WFPC2 images does not appear to include any resolved galaxies.

For NGC~891, we downloaded WFPC2 images from the $HST$ archive and
%obtained 
Advanced Camera for Surveys (ACS) images from the Hubble Legacy
Archive (HLA)\footnote{Based on observations made with the NASA/ESA
{\it Hubble Space Telescope}, obtained from the Hubble Legacy Archive,
which is a collaboration between the Space Telescope Science Institute
(STScI/NASA), the Space Telescope European Coordinating Facility
(ST-ECF/ESA) and the Canadian Astronomy Data Centre (CADC/NRC/CSA).}.
The WFPC2 images we used were: one pointing in F814W at 0.2$\arcm$
from the galaxy center ($HST$ program GO.9042, PI: Smartt), a pointing
in F450W at $r$$\sim$3.5$\arcm$ ($HST$ program GO.8805, PI:
Casertano), and a pointing at $r$$\sim$8$\arcm$ observed in F450W and
F606W ($HST$ program 9676, PI: Rhoads). These WFPC2 data were
processed and analyzed in the same manner as described above.
%The same steps described
%above (``On-the-fly'' calibration, \texttt{gcombine} to create stacked
%images, and photometry with small and large apertures) were used to
%process the WFPC2 images of NGC~4013.  
Thirteen of the WIYN GC candidates were located in one or more of the
WFPC2 frames, and photometry with 0.5-pixel and 3-pixel apertures
indicated that one candidate was a galaxy.
%one of those proved to be a galaxy, based on its
%large count ratio.  
The ACS images we analyzed consisted of three pointings in the halo of
NGC~891 that overlap the region covered by the WIYN pointing. All
three pointings were taken in both the F606W and F814W filters and all
were from $HST$ program GO.9414 (PI: de~Grijs).  The pointings were
located 2.1$\arcm$, 3.7$\arcm$, and 6.8$\arcm$ from the galaxy center.
Images from the HLA are processed using the standard $HST$ pipeline
and combined using the \texttt{Multidrizzle} software.  We measured
the fluxes in two different apertures for 21 of the WIYN GC candidates
in the ACS images and found that three of these were galaxies and ten
were foreground stars.  As a result of this analysis, we decided to
constrain the GC candidate selection applied to the WIYN data by using
a 2-$\sigma$ color selection and by explicitly excluding some of the
background galaxies or stars that remained even after the 2-$\sigma$
criterion was applied.

Our ability to draw general conclusions about the contamination level
in the WIYN data was hampered by the small numbers of objects we could
identify and measure in the $HST$ images.  Therefore, rather than
using the statistics from the $HST$ analysis (e.g., one galaxy out of
13 GC candidates in the WFPC2 images of NGC~891) to apply some sort of
global contamination correction, we opted to simply remove the
extended objects from the WIYN GC candidate list and use another
method to calculate an overall contamination correction.

\subsubsection{Contamination Correction Based on the Asymptotic
Behavior of the 
%GC System 
Radial Profile}
\label{section:asymptotic}

We constructed an initial radial profile for each galaxy's GC system
by assigning the GC candidates to 1$\arcm$-wide concentric annuli centered on
the galaxy nucleus.  We computed the effective area of each annulus 
% in which GCs could be detected 
(the total area minus the regions of the images that were excluded
from the \texttt{daofind} search described in
Section~\ref{section:detection}) and used that
%and the number of GCs
%in the annulus 
to calculate the surface density of GCs for the annulus. GC system
radial distributions that are created from wide-field images in this
manner typically exhibit a peak surface density of GCs near the center
of the galaxy and then fall off monotonically with increasing radius.
At some distance from the galaxy center, they then flatten out to a
fairly constant surface density, as long as the images are large
enough to have covered the full radial range of the GC system (e.g.,
Harris et al. 1985, Harris 1986, RZ01, RZ04, R07).  The constant
surface density in the outermost annuli can be assumed to be due to
contaminating objects, i.e., point sources with magnitudes and colors
like GCs that cannot be easily differentiated from real GCs in a
photometric study.  Thus the surface density in the outer bins can be
subtracted from the overall radial distribution to correct it for
contamination.

%The above-described process was complicated for both NGC~4013 and
%NGC~891.  The initial radial distribution for NGC~4013's GC system did
%show a monotonic decrease and flattening in the outer annuli, but

There were factors that complicated 
%this step in the analysis /
applying this type of asymptotic correction for both NGC~4013 and
NGC~891.  NGC~4013 has recently been shown to have a tidal stellar
stream \citep{martinez09},
%. The stream 
which appears as a low-surface-brightness loop extending
$\sim$6$\arcm$ ($\sim$26~kpc) out from and around the northeast side
of the galaxy.  
%There is also enhanced emission from the disk on the
%southwest side of NGC~4013.  
\citet{martinez09} analyzed the location of the stream relative to
orbital model predictions and concluded that it
%the stream 
was likely due to a dwarf satellite of mass $\sim$6~x~10$^8$ solar
masses that is merging with NGC~4013.  The highest-surface-brightness
portion of the loop appears in our Minimosaic images on the northeast
side of the disk. There may be a slightly enhanced GC population on
that side of the galaxy: when one compares the location of the stream
relative to the GC candidates we detect, it does appear that the
candidates might be preferentially located within the stream, although
the effect is very subtle and may not be real (see
Section~\ref{section:stream}).
%apparently not statistically significant 
%There is some precedent for globular clusters 
%There is of course a confirmed case of a dwarf galaxy 
It does seem possible that the dwarf galaxy responsible for NGC~4013's
stellar stream could have GCs associated with it, based on our current
picture of the best-known accreting dwarf galaxy, the Sagittarius
dwarf spheroidal galaxy (Sgr dSph).  Sgr dSph is being tidally
disrupted by its interaction with the Milky Way and
%theoretical and observational 
studies of its orbit and stellar populations indicate that its mass is
$\sim$10$^9$~M$\sun$ and 
%8 GCs have so far been linked to it
%as many as 
it may have deposited as many as eight GCs into the Milky Way GC system 
%8 GCs may have been associated with it 
(Ibata, Gilmore, \& Irwin 1994; Johnston et al.\ 1999; Layden \&
Sarajedini 2000; Cohen 2004; Carraro et al.\ 2007; Belokurov et al.\
2007; Carraro 2009).
%
%The Sagittarius Dwarf Spheroidal (Sgr dSph) 
%
%Sgr dominant stellar pop is ~6 Gyr (Layden & Sarajedini 2000)
%
%Discovery paper Ibata, Gilmore, & Irwin (1994)
%Galactocentric distance 16 kpc, extending over 3 kpc
%
%Sgr dSph mass = 
%GCs (confirmed) = M54 (postulated to be original nucleus), Arp 2,
%Terzan 7, Terzan 8 (Layden & Sarajedini 2000)
%
%(semi-confirmed) Pal 12 (Cohen 2004)
%Whiting 1 (Carraro et al. 2007)
%
%GCs (possible) = Segue 1 Belokurov et al. 2007
%perhaps even AM 4 Carraro 2009 (not conclusive at all)
%
%Johnston et al. 1999
%best-fit models say current mass is 10^9 solar and it has orbited MW
%for at least 1 Gyr and reduced its mass over that period by factor of
%2-3
%
%SO conclusion is, mass of ~10^9 Msun, and ~8 GCs
%
In order to avoid over-correcting for contamination (and thus
under-counting GCs) in NGC~4013's GC system, we therefore excluded
%from the asymptotic contamination correction 
10 GC candidates that might be associated with the stream from the
asymptotic contamination correction.
% identified 5 GC
%candidates in the annulus centered at 3.8$\arcm$ and 5 candidates in
%the 4.7$\arcm$ annulus as possibly being associated with the stellar
%stream around the galaxy.  We excluded these objects to calculate the
%asymptotic contamination correction for NGC~4013; w
With these objects excluded,
% from the outer two (of five) annuli,
% (and 16 objects remaining), 
the radial profile of GC candidates is fairly constant in the outer
two (of five) annuli and the mean surface density in these annuli is
0.38$\pm$0.09 arcmin$^{-2}$.  (Leaving in the 10 GC candidates leads
to a mean surface density in the outer two annuli of 0.62$\pm$0.12
arcmin$^{-2}$.  Note also that the 10 GC candidates that might coincide
with the stream were removed {\it only} for the purpose of calculating
an asymptotic contamination correction and were included in all
subsequent analysis steps.)

NGC~891 posed a problem for other reasons.  The initial radial profile
created by assigning the GC candidates to 1-$\arcm$ bins did not have
the usual shape 
%expected 
for a GC system, but instead was relatively flat
%(with only small statistical fluctuations) 
over the entire radial extent of the WIYN data. The surface density of
GC candidates fluctuated between 0.41$-$0.96 arcmin$^{-2}$ in the
inner five annuli (at $r$$\sim$1$\arcm$$-$5$\arcm$ from the galaxy
center), before settling down to values between 0.27$-$0.40
arcmin${-2}$ in the last three annuli (at
$r$$\sim$6$\arcm$$-$8$\arcm$).
%It was not clear whether the GC system was even detected (and
%therefore where it ended) so it was difficult to know which annuli to
%use to calculate an asymptotic contamination correction.
We assumed that the last three annuli provided a reasonable estimate
of the surface density of contaminating objects, so we took the
asymptotic correction to be the mean of these values: 0.33$\pm$0.04
arcmin$^{-2}$.

The asymptotic 
%contamination corrections 
values derived from the initial radial profiles were adopted as the
final contamination corrections for 
%the GC systems of 
NGC~4013 and NGC~891.  We note that the stellar contamination levels
estimated from the model (Section~\ref{section:mendez model}) were in
both cases lower than the asymptotic contamination corrections, which
is consistent with the idea that additional contamination comes from
unresolved galaxies.
% with $BVR$ magnitudes like GCs.
A contamination fraction 
%(number of contaminants divided by the
%total number of GC candidates) 
was calculated for each annulus in the radial profile by multiplying
the surface density of contaminants by the effective area of the
annulus and then dividing that number by the total number of GC
candidates in the annulus.  This radially-dependent contamination
correction was later used to correct
%applied to
the GCLF
%globular cluster luminosity function 
(Section~\ref{section:gclf}) and the final radial profile
(Section~\ref{section:radial profile}).

\subsection{Determining the GCLF Coverage of the WIYN Data}
\label{section:gclf}

Observed GC luminosity functions were calculated for NGC~891 and
NGC~4013 by assigning the $V$ magnitudes of the final GC candidates to
bins 0.3 magnitude wide.  A contamination correction was applied to
the data by multiplying the number of GCs in a particular radial bin
by the contamination fraction at that radius, as determined from the
asymptotic correction.  The numbers in each bin were then divided by
the total completeness correction for that magnitude (the total
completeness is calculated by convolving the individual completenesses
in the $BVR$ filters, as described in detail in 
%Section 3.7 of
RZ01). The corrected GC luminosity function for each galaxy was then
fitted with a Gaussian function with a peak absolute magnitude like
that of the Milky Way GCLF, at $M_V$$=$$-$7.3 \citep{az98},
corresponding to an apparent magnitude of $V$$=$ 22.3 and 23.6 for
NGC~891 and NGC~4013, respectively. We varied the dispersion of the
Gaussian between 1.2 and 1.4 mag and found that the mean fraction of
the theoretical GCLF covered by the observed luminosity function was
0.69$\pm$0.01 for NGC~891 and 0.656$\pm$0.002 for NGC~4013. Changing
the bin size of the LF data, or excluding one bin from the fit,
changed the mean fractional coverage by $\sim$2$-$6\%; we included
this uncertainty in the error calculation in Section~\ref{section:spec
freq}.

\subsection{Combining HST Results from the Literature with the WIYN Results}
\label{section:hst2}

The GC systems of the two galaxies we observed with WIYN have been
studied using $HST$ data: Goudfrooij et al.\ (2003; hereafter G03)
analyzed NGC~4013 and Harris et al.\ (2009; hereafter H09) analyzed
NGC~891.  The $HST$ studies are complementary to our analysis:  they
%provide an independent check on the WIYN results and 
have smaller radial coverage but higher resolution than the
ground-based data and allow a search for GCs further into the galaxy
disks.
%(so they allow a search for GCs further into the galaxy disks).  
We used the results from these studies in combination with the WIYN
data
%we combined these data with the WIYN results
to produce a radial distribution for each galaxy's GC system.  We
include the $HST$ results in all subsequent analysis steps.
%sounds awkward but just keep going

G03 analyzed $HST$ WFPC2 images of NGC~4013 and six other spiral
galaxies. The images of NGC~4013 were taken in the F555W and F814W
filters and cover the central $\sim$1.5$-$2$\arcm$ around the galaxy.
The WFPC2 field-of-view is plotted in Figure~\ref{fig:n4013 pointing},
along with the location of the WIYN pointing.
%figured this from wiyn/hst pointing mask -- largest x extent (which
%is roughly radial extent) is 112'' = 1.9'.  Also STScI archive says
%the angular separation of these images is 0.7'.
G03 used steps similar to those we executed on the WIYN images to
detect and select point-source GC candidates by their $V$ magnitudes
and $V-I$ colors.  They published a list of the 50 brightest GC
candidates in NGC~4013; these have $V$ $=$ 21.8 to 24.4.  The $V$=24.4
faint limit matches the limit we imposed on the WIYN data. We compared
each object in the G03 list to the WIYN GC candidate list, to examine
in detail the differences in GC selection in the two data sets.
Twenty-five of the 50 objects in the G03 list are located in the
galaxy disk, in areas of the WIYN images that were masked out because
we could not reliably detect GCs there.  Another 20 of the 50 GC
candidates appear in both the WIYN list and the G03 list.  Of the five
remaining from the G03 list, three barely appear in our WIYN images
and fall below our source detection limits. Two others appear in the
WIYN images but were eliminated as GC candidates because they had
large FWHM values or $BVR$ colors outside the expected region.
%Many of the GC candidates appear in both lists, and
%the few differences between the lists can be ascribed to sensible
%reasons (e.g., faint objects that look ``fuzzy'' in the ground-based
%images and thus were excluded, or WIYN candidates near the $V$ limit
%in the $HST$ images that didn't make it into their list of the 50
%brightest candidates).  
We retained all 50 GC candidates in the G03 list to use in the
subsequent analysis steps. 
%The $HST$ GC candidate list extends further
%into the galaxy disk than the WIYN list, and this yields an extra
%$\sim$20 GC candidates in that inner region.
%so includes an extra $\sim$20 candidates.
% that are in areas of the galaxy that are masked out of the WIYN images.

We assumed that the contamination level in the $HST$ candidate list
was negligible; G03 used detection software that characterized the
sharpness and roundness of a source to remove extended objects, and
for magnitudes brighter than $V$$\sim$24 the contamination from
unresolved background galaxies in these types of $HST$ images should
be very low 
%minimal
(Kundu et al.\ 1999).  To estimate the GCLF coverage of the $HST$
data, we plotted the completeness-corrected luminosity function of the
GC candidates and fitted it with Gaussians of varying dispersions, in
the same manner as for the WIYN sample.  The mean fractional coverage
(averaged for three different Gaussian dispersions) is 0.75$\pm$0.01.
The completeness curves were taken from G03; the completeness
correction was minimal, because the $HST$ data are $\sim$90$-$100\%
complete to $V$$=$24.4.

H09 used deep $HST$ ACS data to identify GC candidates in three fields
(which they designated H1, H2, H3) covering part of the disk and inner
halo of NGC~891. The locations of the H09 ACS fields, relative to the
WIYN pointing, are shown in Figure~\ref{fig:n891 pointing}.  Each
field was imaged in the F606W and F814W filters and H09 selected a
sample of possible GC candidates via visual inspection.  They refined
the sample based on the size, shape, ellipticity, magnitude, and color
of each candidate; their final published list of objects consists of
16 ``Best'' GC candidates and 27 more
%that are less certain and may be contaminants.
that may be GCs or contaminants.  
%The best 16 GC candidates have
%$V$$\sim$20.3$-$23.7 and the total sample of 43 objects has
%$V$$\sim$20.3$-$25.8. 
To simplify the process of combining the $HST$ and WIYN results, we
opted to include in our analysis only the sources in the H1 and H2
fields from H09.  The H3 field covered an area beyond the boundaries
of the WIYN images and H09 identified only five candidates in the
field: two ``best'' GC candidates and three they tentatively label as
galaxies; we did nothing more with the H3 pointing or sources.
%included only five identified candidates

%the lists. 
%to the list from the
%WIYN data and for each source we investigated whether it appeared in
%both lists and if it did not, why it was missing from either list.
%As with the comparison of the G03 data to our WIYN data, 
%In every instance there was a logical explanation for why a given
%source did not appear in both lists (e.g., some $HST$ sources were too
%faint or were located in masked regions in the WIYN images).  
We examined each object in the H09 list and created a final list of GC
candidates from the H09 study to include in the combined ($HST$ plus
WIYN) radial profile.  The final $HST$ GC candidate list has 21
objects and includes the ``Best'' candidates from H09, plus several
objects that were marked ``Bulge'' or ``Disc'' by H09 (to denote that
they might be massive open clusters in the Bulge or Disk of NGC~891).
None of the objects that were noted as possible contaminants by H09
(e.g., designated ``Star?'', ``Galaxy?'', or ``LSB'') survived the
source-by-source review process; in some cases, the objects were
clearly galaxies, even in the ground-based WIYN images.  As part of
this same process, we also compared the lists of GC candidates from
H09 and the WIYN data and investigated the differences between them.
There were very few overlaps between the WIYN and HST GC candidate
lists.  Fifteen of the 21 H09 GC candidates were either in regions of
the WIYN images that had been masked out (regions in the spiral disk
where point sources could not be reliably detected) or located off the
edge of the WIYN images.  Four objects were already identified as WIYN
GC candidates.  Another object had been excluded from the WIYN GC
candidate list because it was bright (V $=$ 19.3); we decided to add
it back to the WIYN list since it was so clearly a GC candidate in the
$HST$ data.  Finally, one object had been excluded from the WIYN GC
candidate list because it is extended in the WIYN images.

The contamination level in the final sample of 21 GC candidates from
H09 should be minimal;
%low/minimal/nominal/negligible/insignificant.  
NGC~891 is only $\sim$8~Mpc away so GCs are sometimes resolved in ACS
images and can usually be distinguished from point-source foreground
stars. H09 make use of this and shape information to remove stars and
galaxies from the sample.  The H09 data are very deep, reaching 50\%
completeness at $V$$\sim$29 and $I$$\sim$27$-$28, whereas the
theoretical GCLF for this galaxy should peak at $V$$\sim$22.3 and go
to zero at $V$$\sim$26.  Because of this, we assumed that the GCLF was
fully sampled by the H09 data and that any GCLF coverage correction
would be zero or negligible.

\section{Results}
\label{section:results}

\subsection{Radial Distributions of the GC Systems}
\label{section:radial profile}

The final, corrected radial distributions of each galaxy's GC system
were produced first by assigning the GC candidates from WIYN and the
HST studies of G03 or H09 to one of a series of concentric annuli,
according to their projected radial distances from the galaxy
center. For the WIYN data, we used annuli of width 1$\arcm$ (as
explained in Section~\ref{section:asymptotic}).  The HST GC candidates
were more densely packed and located within a smaller region, so were
binned more finely, into annuli of width 0.5$\arcm$.  The inner
boundary of the first annulus was always placed just inside the radial
position of the innermost GC candidate.  An effective area, that
excludes missing or masked regions, was calculated for each
annulus. The number of GCs in each annulus was corrected for
contamination and GCLF coverage when appropriate, using the
corrections discussed in previous sections. We then divided the
corrected number of GCs
%in each annulus 
by the appropriate effective area to yield the surface density of GCs
in each annulus. The error on the surface density includes Poisson
errors on the number of GCs and the number of contaminating objects.
The final radial distributions for the GC systems of NGC~891 and
NGC~4013 are listed in Tables~\ref{table:profile n891} and
~\ref{table:profile n4013} and plotted in Figures~\ref{fig:profile
n891} and \ref{fig:profile n4013}. The radii listed in the tables are
the average projected radii of the unmasked pixels in the
corresponding annulus.  For each projected radius, the tables list the
surface density and error, the fraction of the annulus that was
observed (excluding missing or masked regions), and the source of the
measurement (whether it came from the WIYN or HST data).

We note here that neither of the published HST studies of these
galaxies included a calculation of the GC system radial profile. G03
stated that low number statistics and missing spatial coverage made it
too difficult to evaluate the radial distribution of GCs, and they
estimated the total number of GCs instead by comparing the number of
GCs detected at various spatial locations around NGC~4013 with the
number that would be detected at that same location in the Milky
Way. They note that this type of analysis implicitly assumes that the
spatial distribution of GCs in the target galaxy is similar to that of
the Milky Way. H09 comment on the ``severely limited area coverage''
of the ACS study and similarly choose to analyze the spatial
distribution and population of NGC~891's GC system by comparing the
number of GC candidates in their fields with the number that would
fall within the same projected area around the Milky Way.
%Both of those studies estimate a specific frequency for the GC system
%by 
By 
%doing what we have done here and
combining the data from HST that covers the inner few arc~minutes of
the galaxy with the ground-based WIYN images that extend to larger
radii, we are able to quantify the radial profile of the GC systems
and empirically determine how far the systems extend from the host
galaxies.  This leads to better-determined values of the total number
and specific frequency of GCs for each system (see the next section). 
%, because we can fit the radial profile and
%then integrate it over the observed radial extent.

\begin{figure}
%\plotone{profile_n891_for_paper.ps}
\plotone{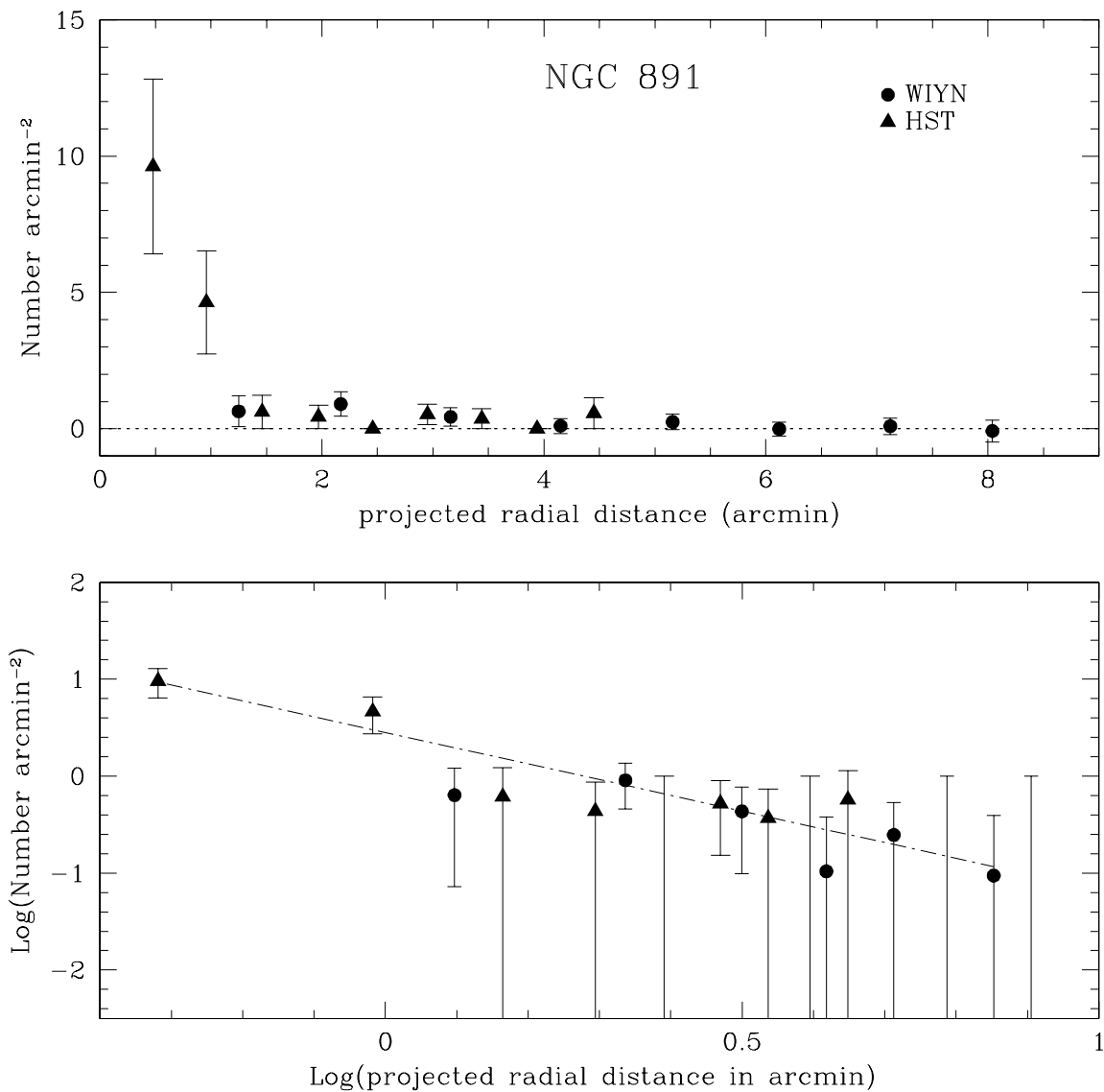}
\caption{\normalsize Radial distribution of GCs in NGC~891, plotted as
  surface density vs.\ projected radial distance, $r$ (top), and as
  the log of the surface density vs.\ log of $r$ (bottom).  The
  horizontal dotted line in the top panel marks zero surface density.
  The dashed line in the bottom panel is the best-fit power law.  The
  data have been corrected for contamination, areal coverage, and
  magnitude incompleteness, as described in
  Section~\ref{section:radial profile}.}
\label{fig:profile n891}
\end{figure}

\begin{figure}
%\plotone{profile_n4013_for_paper.ps}
\plotone{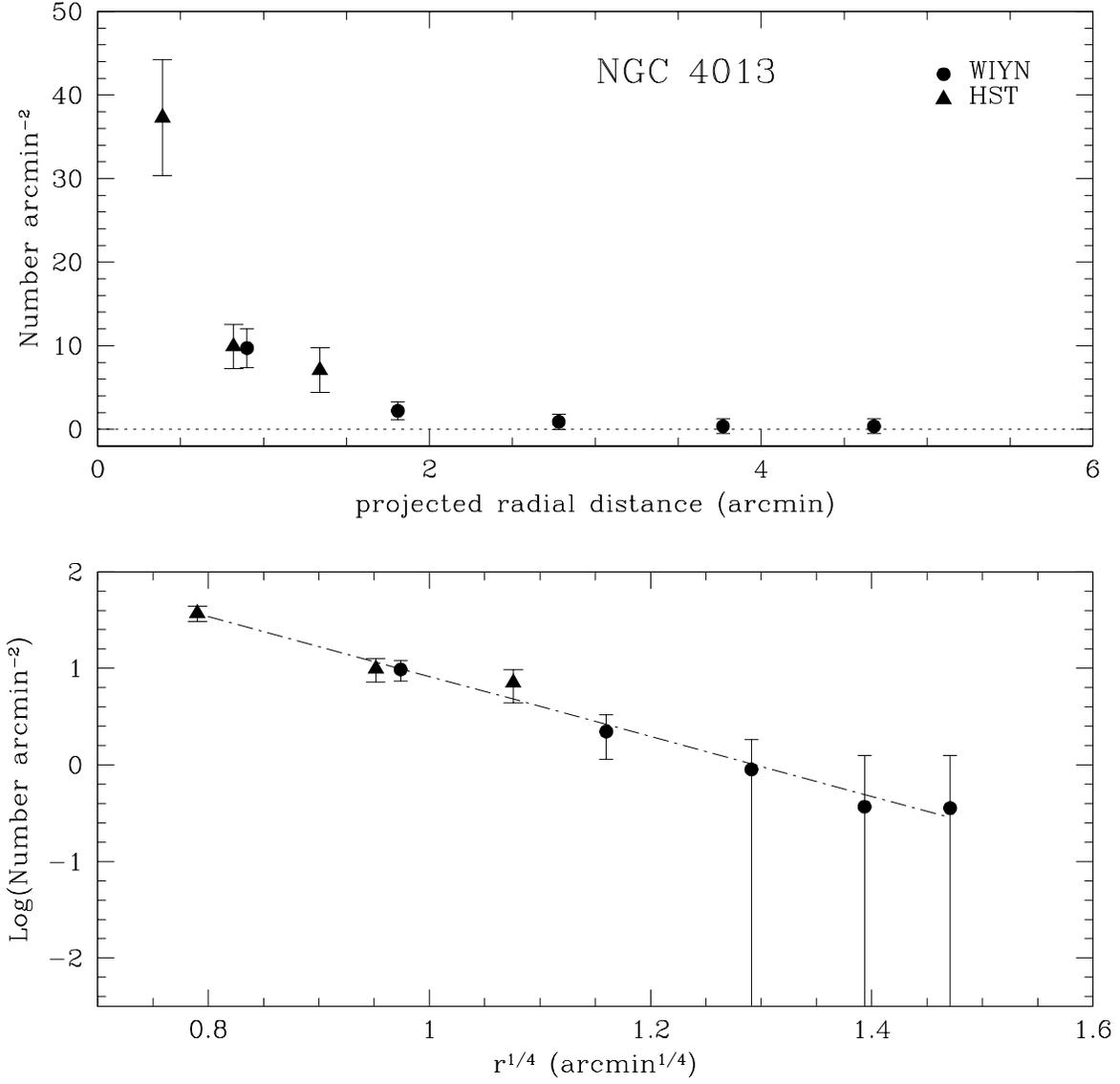}
\caption{\normalsize Radial distribution of GCs in NGC~4013, plotted
  in the same way as in Figure ~\ref{fig:profile n891}. The dashed line
  in the bottom panel is the best-fit de~Vaucouleurs law.}
%as surface density vs.\ projected radial distance, $r$ (top), and as the
%  log of the surface density vs.\ $r^{1/4}$ (bottom).  The horizontal
%  dotted line in the top panel marks zero surface density.   The
%  data have been corrected for contamination, areal coverage, and
%  magnitude incompleteness, as described in
%  Section~\ref{section:radial profile}.}
\label{fig:profile n4013}
\end{figure}

The radial distribution of the NGC~4013 GC system
(Figure~\ref{fig:profile n4013}) has the well-behaved, regular
appearance typical of the GC systems of most of the other spiral
galaxies in our survey (see RZ03 and R07).  The surface density of GCs
is at its highest value near the galaxy center and then decreases
monotonically until it eventually reaches zero within the errors, in
this case in the $r$$=$2.8$\arcm$ ($\sim$12 kpc) annulus. Conversely,
the profile of NGC~891's GC system is not well-behaved.  The combined
$HST$ plus WIYN profile (Figure~\ref{fig:profile n891}) shows positive
surface density in the first two annuli (from the $HST$ data) inside
1$\arcm$, but then the surface density plunges in the third annulus at
1.25$\arcm$, then fluctuates between zero and positive surface density
at $r$$\sim$1.5$\arcm$$-$3.4$\arcm$.  All the annuli beyond
$r$$\sim$3.4$\arcm$ ($\sim$8~kpc) have zero surface density within the
errors. Without the H09 GC candidates to add to the WIYN sample, we
would likely have concluded that NGC~891's GC system was not
definitively detected in the WIYN images.  (The $BVR$ color-color plot
in Figure~\ref{fig:bvr n891} and the color-magnitude diagram in
Figure~\ref{fig:cmds} likewise indicate, at best, a weak WIYN
detection of the GC system: there is little or no overdensity of
objects within the GC selection box in the color-color diagram, and
the objects in the color-magnitude diagram are not tightly clustered
in the same region of the $V$-($B-R$) plane, as they should be.)
Likewise, with only the $HST$ data we would not have known how far out
to integrate the GC system profile in Section~\ref{section:spec freq}.
%radial profilenot have been able to
%trace the radial profile over several arc~minutes, so the shape of the
%profile at large radius would have been uncertain.
%over such a large radial range, so how far
%out to integrate the GC surface density would have been uncertain.

The corrected radial distributions were fitted with de~Vaucouleurs
profiles (of the form log~$\sigma_{\rm GC}$ $=$ $a0$ $+$
$a1$~$r^{1/4}$) and power law profiles (of the form log~$\sigma_{\rm
GC}$ $=$ $a0$ $+$ $a1$~log~$r$).  For NGC~891, the power law had a
slightly smaller $\chi^2$ value (0.9 vs.\ 1.1 for the de~Vaucouleurs
law); for NGC~4013, the opposite was true, and the de~Vaucouleurs
$\chi^2$ was smaller (0.4 vs.\ 0.5 for the power law). The best-fit
coefficients for both the de~Vaucouleurs law and power law, for both
galaxies' GC systems, are given in Table~\ref{table:coefficients}. The
function with the smallest $\chi^2$ is plotted in the bottom panels of
Figures~\ref{fig:profile n891} and \ref{fig:profile n4013} and used to
calculate the total number of GCs in Section~\ref{section:spec freq}.

\begin{figure}
%\plotone{spatial_extent_for_two_spirals_paper.ps}
\plotone{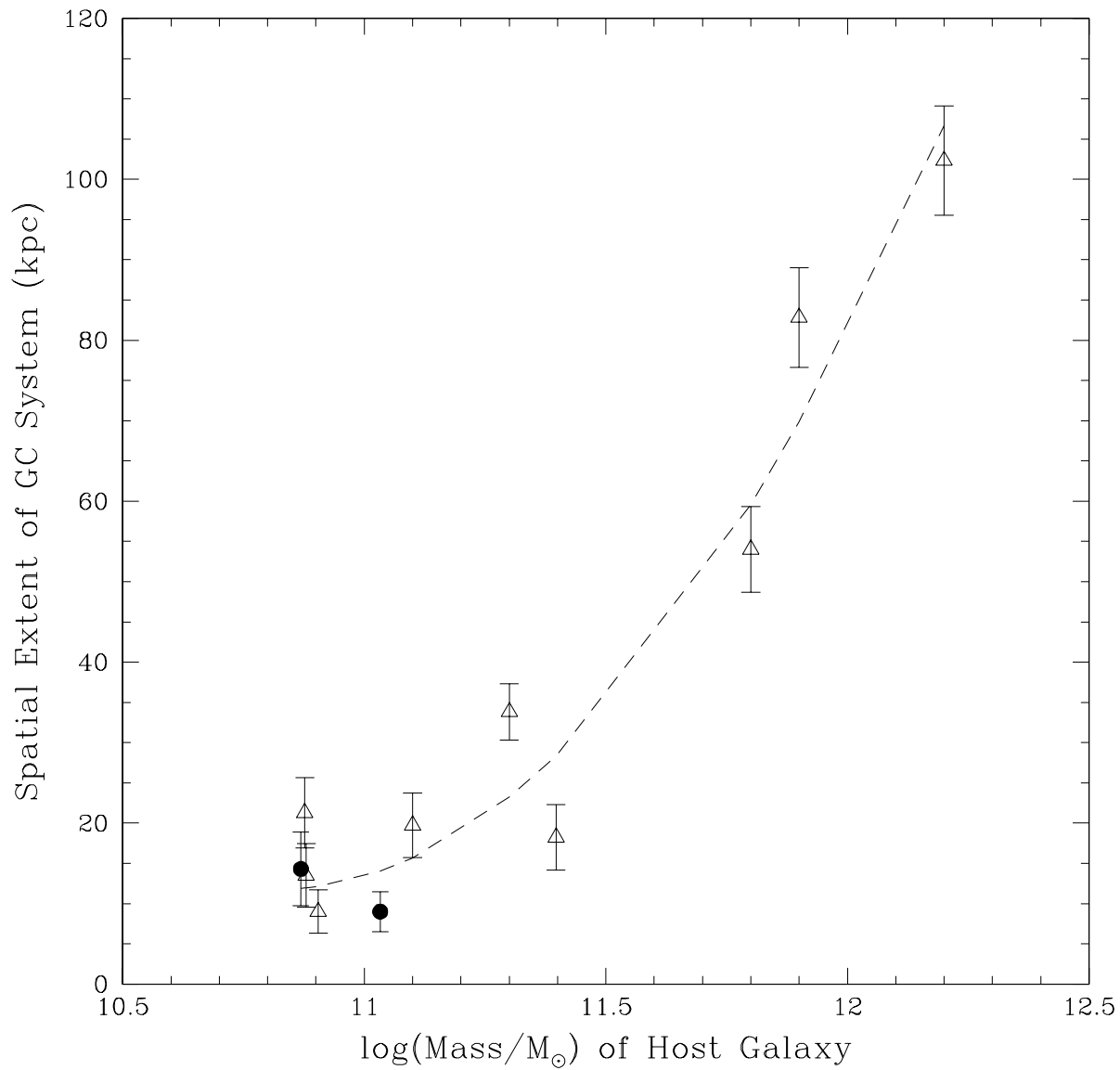}
\caption{\normalsize Radial extent of the GC system in kiloparsecs
  versus the log of the galaxy stellar mass in solar masses for 11
  elliptical, S0, and spiral galaxies included in our wide-field GC
  system survey to date.  Open triangles are the points from R07;
  filled circles are two new points from the current paper.  The
  best-fit second-order polynomial (which is the same, within the errors,
  as the one given in R07) is shown as a dashed line.}
\label{fig:extent}
\end{figure}

One of the quantities we evaluate in the wide-field survey is the
radial extent of each galaxy's GC system, which we define as the point
in the radial profile at which the surface density of GCs becomes
consistent with zero and remains zero (within the errors) to the outer
limit of the data (R07).  We tabulated this measure of radial extent
for nine survey galaxies in R07 and we can now add two more points to
the sample.
% derived the radial extent of nine galaxies studied as part of our
% survey and we can add two more points to this sample.  
The radial extent for NGC~891 is 9$\pm$3~kpc and for NGC~4013 is
14$\pm$5~kpc; the errors include distance errors and uncertainties
associated with determining the extent from binned radial profile
data.
% that have been binned 
%the determination of the extent 
The galaxy stellar masses, derived by combining $M^T_V$
(Table~\ref{table:properties}) with mass-to-light ratios from
\citet{za93}, are log($M$/$M_{\sun}$)$=$11.0 and 10.9 for NGC~891 and
NGC~4013, respectively.  In R07, we showed that the more massive
galaxies in the survey generally have more extended GC systems and we
derived a quantitative relationship between the GC system radial
extent and the host galaxy mass.  As in R07, we again fit a line and a
second-order polynomial to the data.  The best-fit coefficients of both
the line and the curve are consistent within the errors with the
corresponding coefficients derived in R07.  The data and new best-fit
second-order curve are plotted in Figure~\ref{fig:extent}.  The curve has
the form

\begin{equation}
y = ((49.9\pm9.2)~x^2) - ((1080\pm209)~x) + (5860\pm1190),
\end{equation}

\noindent where $x$ is log($M$/$M\sun$) and $y$ is the radial extent
in kiloparsecs.  This relationship can be used to determine in advance
how much spatial coverage is needed 
%in order
%what size detector (i.e., in terms of spatial coverage on the sky) is
%needed 
to image the GC system of a given galaxy.

\subsection{GCs and the Stellar Stream of NGC~4013}
\label{section:stream}

As mentioned in Section~\ref{section:asymptotic}, NGC~4013 was found
by \citet{martinez09} to be surrounded by a low-surface-brightness
stellar stream, possibly due to a dwarf galaxy merging into the larger
galaxy's halo. In the process of calculating the asymptotic
contamination correction for the GC system of NGC~4013, we identified
10 GC candidates that coincide spatially with the stellar stream and
therefore 
%might be associated with it
%might be associated spatially with the stellar stream 
%(i.e., they 
might have originated in the dwarf galaxy that is being tidally
disrupted.  Without velocity measurements for these objects (and
perhaps even then), we cannot definitively say whether or not the GC
candidates are actually moving with the stream material. Since only
three of the relevant GC candidates have $V$$\leq$22.0 and the other
seven have 23.4$<$$V$$<$24.3, velocity measurements would be difficult
to obtain for the majority of the 10 candidates, given the limits of
today's spectroscopic facilities.
%capabilities

We explored various methods for investigating, based on the WIYN
images, whether there are indeed 
%an overdensity of 
``stream'' GCs in NGC~4013.
%determining from the images whether there is indeed an
%overdensity of GCs that might be associated with are
%``stream'' GCs in NGC~4013. 
One straightforward approach is to 
%One way to investigate whether there are ``stream'' GCs around
%NGC~4013 is to 
see whether there is an overdensity of GC candidates in the region of
the stream loop on the galaxy's northeast side 
%of the galaxy 
relative to other areas of the GC system.
%The radial profile of
%NGC~4013's GC system does not show 
To do this, we defined a rectangular 
%``stream'' 
region on the WIYN images of
NGC~4013 that isolates the part of the tidal stream that included the
10 GC candidates that are spatially coincident 
% we thought might be
%that appeared to be 
%associated 
with the stream.  We also defined a ``control'' region with
the same total area and mean radial distance from the galaxy center
and therefore that samples 
%sampling 
an analogous part of the galaxy's GC system.  There are 13 GC
candidates located within the stream region and 10 in the control
region; assuming Poisson statistics, 
%(because we are counting objects),
these numbers are equivalent. Therefore we cannot say, based on our
imaging data and a
%type of 
simple number density argument,
%definitely say 
whether or not there are GCs associated with the tidal stream in
NGC~4013.  This is perhaps not surprising, since it has taken years of
%observational and theoretical 
work -- including theoretical modeling, accurate velocity measurements
and detailed stellar population studies -- to show that Sgr dSph in
our own galaxy's halo is host to several GCs that were once thought to
be part of the Milky Way GC system (e.g., Ibata, Gilmore, \& Irwin
1994, Layden \& Sarajedini 2000, Cohen 2004, Carraro et al. 2007).

\subsection{Total Numbers and Specific Frequencies of GCs}
\label{section:spec freq}

\subsubsection{Total Specific Frequency}

We calculated the total number of GCs ($N_{GC}$) in each galaxy's
system by integrating
%, with respect to radius, 
the function that provides the best fit to the final radial
distribution.
% out to an outer radial limit.  
For NGC~891, we integrated the best-fitting power law given in
Table~\ref{table:coefficients}.  The inner radius of the integration
was taken to be 0.2$\arcm$, the inner boundary of the first annulus of
the observed radial profile.
%innermost radius for which we have data out to the 
The outer radius of the integration was 3.7$\arcm$, because beyond
that radius the surface density of GCs is consistent with zero (within
the errors) in the rest of the annuli. We had to make assumptions
about the behavior of the radial profile near the galaxy center, in
regions we could not observe.  For NGC~891, this was the central
0.2$\arcm$ of the GC system, or the central $\sim$490~pc for $m-M$
$=$29.61.  We calculated the number of GCs that had been missed in
this inner region given two possibilities: (1) that the profile inside
the unobserved region continued to follow the best-fit power law to
small $r$, or (2) that the profile flattened in the inner region
(i.e., the GC surface density in the unobserved region
% inside
%0.2$\arcm$ 
equaled the value in the innermost observed annulus). Adding the
number of GCs in the observed portion of the system to the number
given these two assumptions for the inner region yielded two estimates
for $N_{GC}$: 53 for the flat inner profile and 81 if we extended the
power law to small $r$. The mean of the two estimates, $N_{GC}$ $=$
70$\pm$20, is the final value given in Table~\ref{table:total
numbers}.

For NGC~4013, the de~Vaucouleurs law in Table~\ref{table:coefficients}
was the best-fitting function to the observed radial profile.  We
integrated the de~Vaucouleurs law from 0.08$\arcm$ (the inner boundary
of the first annulus in the observed profile) to 3.25$\arcm$ (the
outer boundary of the 2.8$\arcm$ annulus, which is where the GC
surface density reaches zero within the errors and remains zero for
the rest of the observed profile).  For the region inside 0.08$\arcm$,
we assumed a flat radial profile or a de~Vaucouleurs law all the way
to $r$$=$0.  The results given these two assumptions were very close
to each other: $N_{GC}$ $=$ 138 for the flat inner profile and 144 if
the de~Vaucouleurs law was continued to $r$$=$0.  Again we have taken
the average of the two values for $N_{GC}$ to be the final value for
the GC system: 140$\pm$20 (Table~\ref{table:total numbers}).

We also calculated luminosity- and mass-normalized specific
frequencies for each galaxy's GC system.  The total number of GCs
normalized by the $V$-band luminosity of the host galaxy is defined as

\begin{equation}
{S_N \equiv {N_{GC}}10^{+0.4({M_V}+15)}}
\end{equation}

\noindent \citep{hvdb81} and the number of GCs normalized by the host galaxy
stellar mass is defined as

\begin{equation}
T \equiv \frac{N_{GC}}{M_G/10^9\ {\rm M_{\sun}}}
\end{equation}

\noindent \citep{za93}.  We used $M^T_V$ from
Table~\ref{table:properties} and the galaxy mass (designated $M_G$
above) from Section~\ref{section:radial profile} to calculate $S_N$
and $T$; specific frequencies are given in Table~\ref{table:total
numbers}.

The errors on $N_{GC}$ given in Table~\ref{table:total numbers}
include contributions from the following sources of uncertainty: (1)
the change in the calculated GCLF coverage depending on the assumed
intrinsic GCLF dispersion and how the LF data were binned; (2) Poisson
errors on the number of GCs and contaminating objects; and (3)
uncertainties in the number of GCs in the inner, unobserved region of
the system. The errors on the specific frequencies $S_N$ and $T$ also
take into account the error on the total galaxy magnitude; we assumed
that the uncertainty in the extinction-corrected galaxy magnitude was
three times the error on $V^T$ given in RC3 \citep{devauc91}.  Each of
these sources of uncertainty was added in quadrature to yield the
final total error value given in the table.

%As was noted in Section~\ref{section:radial profile}, when G03 used
%$HST$ WFPC2 imaging to study the GC system of NGC~4013, they did not
%construct a radial profile but instead derived the total number 
Next we examine how our estimates of the number and specific frequency
of GCs for these galaxies compare to those from past studies. 
%previous values.  
In their $HST$ WFPC2 study of the GC system of NGC~4013, G03 estimate
the total number of GCs not by constructing and integrating the radial
profile, but by comparing the number of GC candidates they detect in
certain regions around the galaxy to the number of GCs that would be
detected at analogous locations in the Milky Way.  Using this method,
they calculate that NGC~4013 has $N_{GC}$ $=$ 243$\pm$51, which is
significantly larger than our estimate of 140$\pm$20.  They combine
this with $m-M$ $=$ 31.35 and $M_V^T$ $=$ $-$20.83 to derive $S_N$ $=$
1.1$\pm$0.3 and $T$ $=$ 2.2$\pm$0.7.  They adopt a larger distance
modulus than we do for NGC~4013, so their inferred galaxy luminosity
and mass are larger than ours.  Our value for $N_{GC}$ is $\sim$40\%
lower than their value, but our $S_N$ and $T$ values actually agree
with the estimates given in G03 because of the larger distance modulus
they assume.
If we instead assume the distance modulus from G03, $m-M$ $=$ 31.35,
and go through all of the analysis steps to derive final numbers and
specific frequencies for NGC~4013 (namely, applying a new $V$
magnitude selection to the GC sample, fitting the GCLF for both the
WIYN and HST data, and calculating and integrating a new radial
profile), we find $N_{GC}$ $=$160$\pm$20, $S_N$ $=$ 0.7$\pm$0.2, and
$T$ $=$ 1.4$\pm$0.3.  The end result is that, even assuming the larger
distance to NGC~4013, we derive a smaller total number of GCs in the
galaxy which leads to correspondingly smaller $S_N$ and $T$ values.
%
%The bottom line is that, 
%If we used the same value for $M_V$ as G03, this study would find
%$S_N$ and $T$ values that are smaller by about $\sim$40\%.
%The end result is that
%their $S_N$ and $T$ values actually agree with our estimates despite
%the fact that our $N_{GC}$ is $\sim$40\% lower than their value.

The GC system of NGC~891 has been studied by van den Bergh \& Harris
(1982) and much more recently by the H09 $HST$ study that we have
combined with our WIYN results.  van den Bergh \& Harris (1982)
obtained photographic observations with $\sim$1$\arcsec$ seeing and
counted point sources around NGC~891, looking for an overdensity of
point-like objects that would signal the presence of a GC
population. They found ``no significant population'' of GCs in NGC~891
and calculated a specific frequency consistent with zero ($S_N$
$\sim$0.04$\pm$0.08); they speculated that the weak bulge of NGC~891
may be responsible for its apparent lack of GCs.
%The H09 study of NGC~891 that we 
To calculate $N_{GC}$, H09 adopt a method similar to that of G03:
because their $HST$ ACS data also cover only a small portion of the GC
system, they compare the number of GC candidates they find with the
number of Milky Way GCs that would fall within a corresponding
projected area.  From this comparison they conclude that NGC~891 has
between 80 and 200 GCs and that the GC system ``resembles that of the
Milky Way rather closely'' in terms of total number and spatial
distribution of GCs.  By combining the H09 GC candidate list with
%wide-field 
WIYN imaging data and constructing a radial profile for the GC system,
we find that NGC~891 has a fairly small total number of GCs
(70$\pm$20) and small specific frequency ($S_N$ $=$ 0.3$\pm$0.1,
compared to 0.6$\pm$0.1 for the Milky Way).
%  although not outside the
%scatter for spiral galaxies in general.  
%It seems likely that, 
Without the $HST$ results from H09 to augment our WIYN data, we might
have concluded 
%as van den Bergh \& Harris (1982) did 
that NGC~891 had no detectable GC system as van den Bergh \& Harris
(1982) did.  This galaxy's GC system is so poorly populated, and the
few GCs that are present are so close to the galaxy disk, that typical
ground-based capabilities and techniques are simply not adequate to
produce a definitive detection. On the other hand, combining our
ground-based data with the $HST$ data allowed us to draw more
definitive conclusions than H09 about how many GCs are present in
NGC~891.

\begin{figure}
%\plotone{spiral_morph.ps}
\plotone{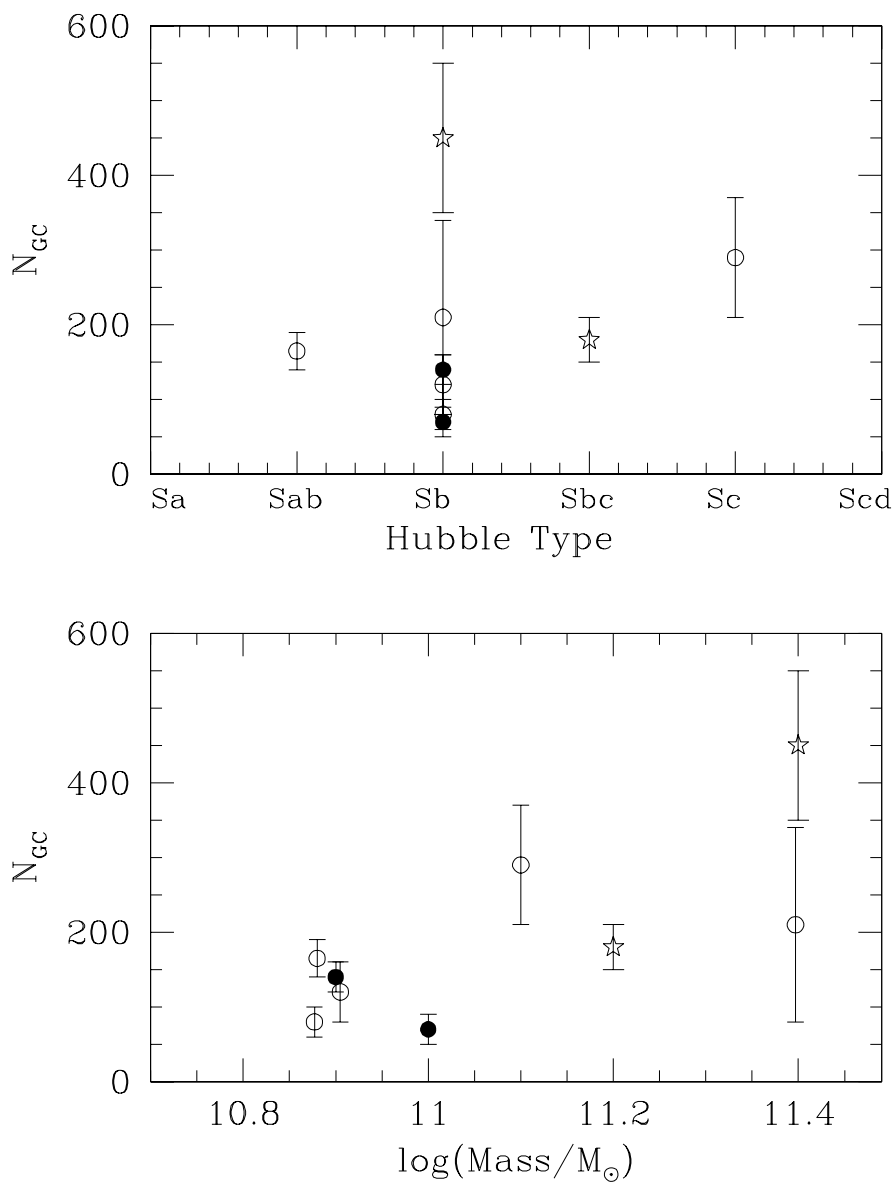}
\caption{\normalsize Number of GCs plotted vs.\ Hubble type (top) and
  galaxy stellar mass (bottom).  Values for the spiral galaxies
  analyzed for our wide-field survey are plotted with filled circles
  (the two galaxies from the current paper) or open circles (four
  galaxies from R07 and one from RZ03).
%seven spiral galaxies
%  analyzed for our wide-field survey (two from the current paper, four
%  from R07, and one from RZ03) are shown as filled circles.  
Values for the Milky Way (Ashman \& Zepf 1998) and M31 (Ashman \& Zepf
  1998, Barmby et al.\ 2000; larger error bars) are plotted with open
  stars.}
\label{fig:spiral morph}
\end{figure}

\begin{figure}
%\plotone{spiral_spec_freq.ps}
\plotone{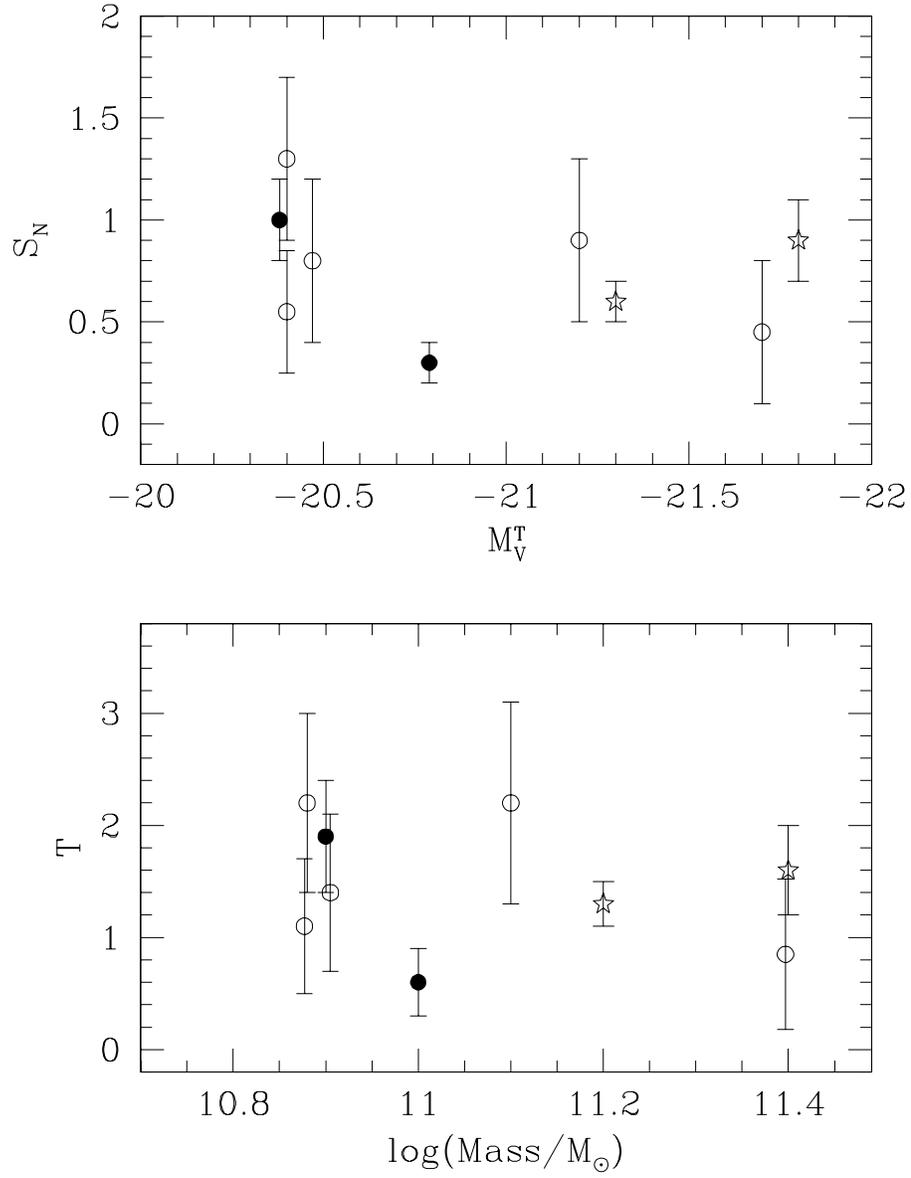}
\caption{\normalsize $S_N$
%Number of GCs normalized by $V$-band galaxy
%  luminosity ($S_N$) 
vs.\ $V$-band galaxy luminosity (top) and 
%number of GCs normalized by galaxy stellar mass ($T$)
$T$ vs.\ galaxy stellar mass (bottom).  Symbols are the same as in
Figure ~\ref{fig:spiral morph}.}
%Values for the seven spiral
%  galaxies analyzed for our wide-field survey (two from the current
%  paper, four from R07, and one from RZ03) are designated by filled
%  circles.  Values for the Milky Way (Ashman \& Zepf 1998) and M31
%  (Ashman \& Zepf 1998, Barmby et al.\ 2000; {\it larger error bars})
%  are plotted with open stars.
\label{fig:spiral spec freq}
\end{figure}

Figure~\ref{fig:spiral morph} plots the number of GCs against galaxy
Hubble type (top panel) and against the log of the galaxy stellar mass
(bottom panel) for the spiral galaxies in our wide-field survey, plus
the Milky Way and M31. Figure~\ref{fig:spiral spec freq} shows $S_N$
versus $M_T^V$ and $T$ versus the log of the galaxy stellar mass for
the same set of galaxies.  In R07, we calculated the mean value of
$N_{GC}$, and the weighted mean values of $S_N$ and $T$, for the
spiral galaxies in our survey plus the Milky Way and M31.  These were
$N_{GC}$ $=$ 170$\pm$40, $S_N$ $=$ 0.8$\pm$0.2, and $T$ $=$
1.4$\pm$0.3.  The values we calculate for NGC~4013's GC system are
very much in line with these mean values,
% are very much in line with these
%mean values for the overall sample, 
but the $N_{GC}$, $S_N$, and $T$ values for NGC~891 are comparatively
low.  NGC~891 is frequently described in the literature as being
similar to the Milky Way because of its morphological type,
luminosity, and molecular gas distribution (e.g., Sandage 1961, van
der Kruit 1984, Scoville et al.\ 1993), but its GC system 
%(we find) 
is about half as populous as the Milky Way's. One possibility we
considered is whether this galaxy has very few GCs because, despite
being similar to the Milky Way in appearance, it is actually much less
massive than our Galaxy. As explained in Section~\ref{section:radial
profile}, we calculate log(M/M$_{\sun}$) $=$ 11.0 for NGC~891.  The
Milky Way has $M_V$ $=$ $-$21.3 and is an Sbc spiral galaxy
\citep{harris91}; assuming the $M/L_V$ values from \citet{za93}, this
yields a mass of log(M/M$_{\sun}$) $=$ 11.2.  The CO $+$ HI rotation
curve of NGC~891 is similar to that of the Milky Way, with a rotation
velocity that rises steeply near the galaxy nucleus to a maximum
velocity of 250~km~s$^{-1}$ at $\sim$3~kpc out to 15~kpc
\citep{sofue96}, suggesting this galaxy and the Milky Way probably
have comparable mass distributions.  In any case, the new mean value
for $N_{GC}$ for spiral galaxies in Figure~\ref{fig:spiral morph}
%our wide-field survey (including NGC~891 and NGC~4013) 
is 150$\pm$30.  The weighted mean values for $S_N$ and $T$ for the
galaxies in Figure~\ref{fig:spiral spec freq} are 0.5$\pm$0.1 and
1.1$\pm$0.2, respectively. These updated values agree within the
errors with results from Chandar et al.\ (2004), who studied five
spiral galaxies and found a mean $S_N$ of 0.5$\pm$0.1 and $T$ of
1.3$\pm$0.2.  However, our new mean values are slightly smaller than
the average values from the G03 study of five Sab$-$Sc spiral
galaxies; G03 found a mean $S_N$ of 0.96$\pm$0.26 and $T$ of
2.0$\pm$0.5.
%
%I'm actually not sure weighted mean is the right approach, b/c the errors
%are propto the numbers of GCs and it's going to weight more heavily the
%smaller values b/c they'll have smaller errors.  So in a sense it
%makes the mean smaller, artificially.  On the other hand, there is
%more than just number of GCs rolled into the error on S and T, so in
%that sense the weighting seems appropriate. 
%
%Weighted mean for N_{GC} is 113 +/- 10; so quite different
%Straight mean for S_N is 0.76 +/- 0.13
%Straight mean for T is 1.46 +/- 0.25
%

The sample of spirals we have observed with the wide-field survey is
  still fairly small, not all morphological types are sampled, and we
  have more Sb spiral galaxies in the survey than any other type.
  Therefore we do not wish to over-interpret the significance of the
  results.  However, based on this small sample, there does seem to be
  modest
%a fair amount of 
variation in the number of GCs and in the GC specific frequency for
  galaxies with similar properties.
%within each spiral galaxy class and even among galaxies
%  with similar masses.  
For example, the Sb galaxy NGC~4157 and the Sab galaxy NGC~7814 have
  almost the same stellar mass (log$(M/M_{\sun}$ $=$ 10.9), but the
  number of GCs (80$\pm$20 and 170$\pm$30, respectively) differs by a
  factor of two.  NGC~891, NGC~2683, NGC~4157, and NGC~4013 are all
  classified as Sb spiral galaxies, but the number of GCs they host
  ranges from 70 to \gapp200.  The bottom line seems to be that in the
  overall sample, there may be a weak trend of increasing $N_{GC}$
  with increasing galaxy mass (bottom panel of Figure~\ref{fig:spiral
  morph}), but there is enough variation in $N_{GC}$ at each galaxy
  mass, magnitude, and/or Hubble type to produce scatter in the $T$
  versus galaxy stellar mass and $S_N$ versus galaxy magnitude
  relationships seen in Figure~\ref{fig:spiral spec freq}.

\subsubsection{Specific Frequency of Blue (Metal-Poor) GCs}

Many giant galaxies -- including the two 
%giant galaxies 
with arguably the most thoroughly studied GC systems, the Milky Way
and M31 -- have been shown to possess GC systems with at least two
populations: a blue, metal-poor population and a redder, more
metal-rich populations
\citep{zinn85,za93,barmby00,kw01,perrett02,kz07,strader07}.
%Spectroscopic metallicities,
%kinematics, and near-IR colors of GCs confirm the interpretation that
%giant galaxies often have two populations of 
Various galaxy formation models have either predicted the presence of
these multiple populations of GCs \citep{az92} or sought to explain
them in the context of the galaxies' assembly histories
\citep{fbg97,cote98,beasley02}.  The metal-poor population of GCs is
of particular interest because its low metallicity implies that it 
% presumably 
represents the first generation of GCs created in the early stages of
the host galaxies' formation histories.
%formation
%histories of the host galaxies (or in their progenitors, if the
%massive galaxies are presumed to have assembled from smaller
%protogalaxies or galaxy fragments).

In previous papers from the wide-field survey (R05, R07), we estimated
the number of blue (metal-poor) GCs normalized by the galaxy mass, or 
%a quantity we call 
$T_{\rm blue}$:

\begin{equation}
T_{\rm blue} \equiv \frac{N_{GC}(\rm blue)}{M_G/10^9\ {\rm M_{\sun}}}
\end{equation}

\noindent Here, $N_{GC}(\rm blue)$ is the number of blue GCs and $M_G$
is the stellar mass of the host galaxy, calculated as described in
Section~\ref{section:radial profile}.
%previous section.  
In R05 and R07, we found that $T_{\rm blue}$ for the spiral galaxies
in the survey is significantly smaller than $T_{\rm blue}$ for the
massive cluster ellipticals.
% As discussed in R05 and R07, 
%and argued by several previous
%studies of the {\it total} number of GCs in spirals and ellipticals
%(e.g., Harris 1981, van den Bergh \& Harris 1982) 
This leads to the conclusion that simply merging the GC systems of two
or more late-type galaxies {\it as we see them today} 
cannot
%is not sufficient to 
account for the comparatively large blue GC populations in some giant
ellipticals.
%; a more complicated scenario is required to
%explain how elliptical galaxies ended up with relatively large
%populations of blue GCs.  
We also found 
%in the survey data 
a rough trend of increasing $T_{\rm blue}$ values with increasing
galaxy mass.
% (R05, R07).

We calculated $T_{\rm blue}$ for NGC~891 and NGC~4013 
%in order 
to see how they compare to the values for the other survey galaxies.
and whether they fit within this overall trend.
%$T_{\rm blue}$ was
%computed using the same general method as was used for the other
%spiral galaxies in the survey.  
We first construct a sample of GC candidates with magnitude
completeness of at least 90\% in all the filters; we refer to this
sample of candidates as the ``complete sample''.  We then determine
the fraction of GC candidates in the complete sample that are bluer
than some pre-determined color. The color criterion we use is based on
where the usual separation between the blue and red GC populations is
located in elliptical galaxies; in $B-R$ this occurs at $\sim$1.23
(RZ01, RZ04) and in $V-I$ it occurs at $\sim$1.0 (Kundu \& Whitmore
2001).
%So in order to
%compare the metal-poor GC populations of a sample of giant galaxies,
%$B-R$$=$1.23 seems as appropriate a criterion as any.

For NGC~891, we used the WIYN data to construct a complete sample of
GC candidates and found that seven of the 26 candidates (27\%) in the
complete sample had $B-R$ bluer than 1.23.  Because we had so few
objects in the complete sample from WIYN, we also decided to construct
a complete sample of GC candidates using the HST list from H09.  The
complete sample from the H09 data included 21 GC candidates; 13 of
these (63\%) had $V-I$ $<$ 1.0.  We multiplied each of these blue
fractions times the $T$ value for the galaxy's GC system to calculate
$T_{\rm blue}$ and then took the mean of the two $T_{\rm blue}$ values
as the final value.  We also assigned a generous error that
incorporates a wide range of possible values for $T_{\rm blue}$, given
that working with very small samples of GC candidates makes the entire
process of the $T_{\rm blue}$ calculation for these spiral galaxies
very uncertain.  The final $T_{\rm blue}$ value for NGC~891 is
0.3$\pm$0.2.  For NGC~4013, the complete sample from the WIYN data
included 51 GC candidates; of these, 38 (75\%) have $B-R$ $<$ 1.23.
Multiplying this fraction times the calculated $T$ value for
NGC~4013's GC system yields a $T_{\rm blue}$ value of
1.4$\pm$0.3. These values are listed, along with the other total
numbers and specific frequencies, in Table~\ref{table:total numbers}.

\begin{figure}
%\plotone{tblue_mass_13oct09.ps}
\plotone{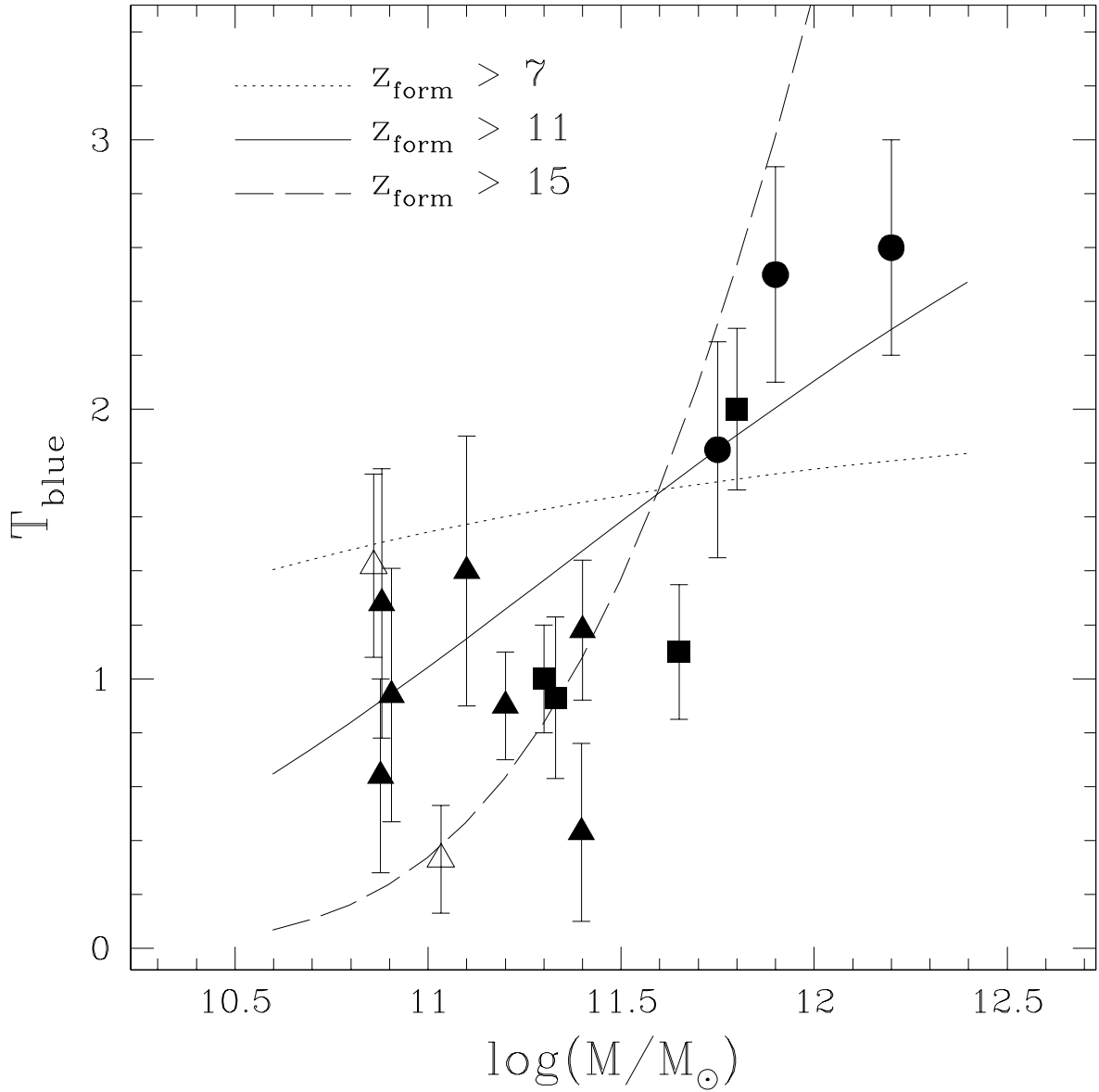}
\caption{\normalsize Galaxy-mass-normalized specific frequency of blue
  (metal-poor) GCs plotted versus the log of the galaxy stellar mass
  for sixteen galaxies from our wide-field survey and from other
  multi-color wide-field CCD imaging studies in the
  literature. Circles represent early-type galaxies in clusters;
  squares indicate early-type field galaxies; and triangles indicate
  field spiral galaxies.  Open symbols denote the two new values from
  the current study.  The curves (courtesy G.\ Bryan) show the slope
  of the $T_{\rm blue}$ vs.\ galaxy stellar mass relation depending on
  whether metal-poor GCs were formed at $z$ $>$ 7, 11, or 15 (see the text
  for details).}
\label{fig:tblue}
\end{figure}

In Figure~\ref{fig:tblue}, we present $T_{\rm blue}$ values for
16 galaxies, plotted versus the galaxies' stellar masses.  Note
that this is the same $T_{\rm blue}$ figure presented in R07, but with
NGC~891 and NGC~4013 added.  The figure now includes: 11 spiral,
S0, and elliptical galaxies from our wide-field survey (RZ01, RZ03,
RZ04, R05, R07); three early-type galaxies from other multi-color,
wide-field CCD studies in the literature (Forbes et al.\ 2001, Gomez
\& Richtler 2004, Harris et al.\ 2004);
%N1052 = E4 = Forbes et al 01
%N5128 = E0 = Harris et al 04
%N4374 = E1 = Gomez & Richtler 04
and the Milky Way and M31 (Zinn 1985, Ashman \& Zepf 1998, Barmby et
al.\ 2000, Perrett et al.\ 2002).  Circles designate giant elliptical
galaxies in clusters, squares are used for early-type galaxies in the
field, and triangles designate spiral galaxies in the field or in
groups.  $T_{\rm blue}$ for the Milky Way
%(Zinn 1985; Ashman \& Zepf 1998) 
and M31 
%(Barmby et al.\ 2000; Perrett et
%al.\ 2002) 
are the triangles with relatively small error bars: $T_{\rm blue}$ $=$
0.9$\pm$0.2 and 1.2$\pm$0.3, respectively.  The $T_{\rm blue}$ value
for NGC~4013 (1.4$\pm$0.3) is very close to the mean derived
%value 
for the other spiral galaxies without the two new points
(1.2$\pm$0.1).
%: the mean $T_{\rm blue}$ for the spiral galaxies
%without the two new points is 1.2$\pm$0.1.  
$T_{\rm blue}$ for NGC~891, however, is relatively low, at only
0.3$\pm$0.2; this is comparable to the value we found for the spiral
galaxy NGC~7331, 0.4$\pm$0.3 (R07). 

The general trend identified in R05 --- i.e., larger numbers of blue,
metal-poor GCs per galaxy stellar mass for more massive galaxies --- 
%higher $T_{\rm blue}$ values for  
is still present in this latest version of the $T_{\rm blue}$ figure,
although (as before) there is a fair amount of scatter in the
relationship. In R05 we suggested that the general trend appears
consistent with a scenario in which galaxies and their GC systems are
formed in a biased, hierarchical assembly process. As explained in
Santos (2003), the idea is that the formation of giant galaxies began
at high redshift ($z$ $>$ 20) as gaseous protogalactic building blocks
collided and merged together to form larger galaxies.
% (the canonical ``hierarchical merger'' picture).  
Merging of gaseous protogalaxies triggers the formation of GCs (in a
similar way as envisioned by Ashman \& Zepf 1992).  As galaxies are
built up hierarchically over time, they increase both their total
stellar mass and their GC population.  Massive galaxies that are
located in parts of the universe that have higher density (e.g., areas
of the universe where galaxy clusters will eventually be located, at
$z$ $\sim$ 0)
%will someday be located) 
begin this assembly process earliest, before less-massive galaxies
that are located in lower-density environs.
%In this way, galaxies build up
%their total stellar masses as well as their GC populations as they
%undergo more and more mergers and accrete more protogalactic 
In the Santos (2003) picture, galaxy assembly and GC formation are
temporarily suppressed when reionization occurs in the universe.  The
energy imparted to the intergalactic medium slows down or stops
protogalactic building blocks from merging and forming stars, perhaps
for a period of \lapp1~Gyr.  Stars continue to evolve during this
time, so that when the assembly and associated GC formation process
resumes, the gas has been enriched and any GCs formed after the
suppression period are metal-rich by comparison.  The result of such a
scenario is 
%would be
that today's more massive galaxies should have higher numbers of
metal-poor GCs per galaxy mass than less massive galaxies, simply
because they started the assembly process first; a larger
percentage of their baryonic mass was in place to participate in the
formation of the first generation of GCs before the temporary
suppression occurred. This type of scenario also provides a natural
explanation for the gap in metallicity between GC subpopulations in
the Milky Way and other giant galaxies.

The curves in Figure~\ref{fig:tblue} were made by G.\ Bryan
(private communication) using a Press-Schecter calculation \citep{ps74}
%that is 
first described in R07.  The curves show how the slope of the
$T_{\rm blue}$ trend would be expected to change as the redshift of
suppression (i.e., the redshift at which galaxy assembly and GC
formation are temporarily stopped) changes, for three choices of
redshift: $z$ $=$ 7, 11, or 15.
% with the three lines
%representing the trend if metal-poor GCs are formed prior to $z$ of 
%
%metal-poor GCs are formed prior to redshift when galaxy assembly
%and GC formation were suppressed: $z$ of 7, 11, or 15.  
Bryan's calculation assumes that GCs form within gaseous building
blocks with masses of at least 10$^8$ solar masses and that the number
of GCs in a given galaxy is proportional to the fraction of the galaxy
mass that has assembled by the suppression redshift.  The current data
do not seem to match the $z$ $=$ 7 curve, but instead suggest a larger
suppression redshift of $z$$\sim$11$-$15.  (Given that Santos (2003)
suggests that reionization is the mechanism that halts the formation
of the first generation of GCs, it is interesting to note that the
$z$$\sim$11$-$15 curve matches well with recent estimates for the
reionization redshift from WMAP (Alvarez et al.\ 2006)).

The biased, hierarchical scenario outlined here is not the only way
that the trend in $T_{\rm blue}$ seen in Figure~\ref{fig:tblue} could
arise; other possible explanations (e.g., differences in the
efficiency of dynamical destruction of GCs in different types of
galaxies, a trend in the relationship between galaxy luminosity and
galaxy stellar mass that would affect the log(M/M$_{\sun}$) values in
the figure) are discussed in R05 and R07.  Our intent here is simply
to show how the $T_{\rm blue}$ and stellar mass values of NGC~891 and
NGC~4013 compare with the values for the other survey galaxies.
%lie in the figure.
%in the $T_{\rm
%  blue}$-galaxy mass relationship.
The figure --- and our understanding of the relationship between
$T_{\rm blue}$ and galaxy and GC system formation and evolution ---
would certainly benefit from having additional well-determined $T_{\rm
blue}$ values at the high-mass end (where we only have values for two
Virgo cluster ellipticals) and in the moderate-luminosity region
(log(M/M$_{\sun}$) $\sim$11.5$-$12).  Accordingly, we have an ongoing
effort to observe more galaxies with multi-color wide-field imaging
and fill in the gap in the relevant mass range.

\section{Summary and Conclusions}
\label{section:summary}

We have combined $BVR$ imaging observations from the
9.6$\arcm$x9.6$\arcm$ Minimosaic imager on the WIYN 3.5-m telescope
with archival and published data from $HST$ to 
%observations to
investigate the GC system properties of two spiral galaxies, NGC~891
and NGC~4013.  The main findings of this work are as follows:

\begin{enumerate}

\item We used both the WIYN data and the $HST$ data to construct
  radial distributions of the GC systems of the galaxies, corrected to
  account for missing areal coverage, the portion of the GCLF we could
  not observe given our detection limits, and contamination from
  foreground stars and background galaxies.  NGC~891's GC system
  radial profile extends to 9$\pm$3~kpc and NGC~4013's extends to
  14$\pm$5~kpc from the centers of the host galaxies; beyond those
  projected radii, the surface density of GCs is consistent with zero
  within the errors.

\item We fitted the radial distributions with de~Vaucouleurs
  and power laws and integrated the best-fitting function to derive
  the total number of GCs.  We find $N_{GC}$ $=$ 70$\pm$20 for NGC~891
  and 140$\pm$20 for NGC~4013. Despite the fact that
  NGC~891 is often considered a Milky Way analog,
% and a previous study of its GC system described it as being ``quite similar
%  ... '' (cite H09), 
we estimate that it has less than half as many GCs as the Galaxy.  Its
  specific frequency $S_N$ is 0.3$\pm$0.1 and its mass-normalized
  number of GCs, $T$, is 0.6$\pm$0.3, compared to $S_N$ $=$ 0.6$\pm$0.1
  and $T$ $=$ 1.3$\pm$0.2 for the Milky Way \citep{az98}.  We find
  that NGC~4013 has $N_{GC}$ $=$ 140$\pm$20, $S_N$ $=$ 1.0$\pm$0.2,
  and $T$ $=$ 1.9$\pm$0.5.

\item We combine the results from this paper with results from our
ongoing wide-field GC system survey and
%of GC systems of giant galaxies and
calculate the average total number and specific frequency of GCs for
spiral galaxies in the survey plus the Milky Way and M31.  We find
$N_{GC}$ $=$ 150$\pm$30, $S_N$ $=$ 0.5$\pm$0.1, and $T$ $=$ 1.1$\pm$0.2.
These values are consistent with values from our most recent survey
paper (R07) and from an analysis of five spiral galaxy GC systems by
Chandar et al.\ (2004) but slightly smaller than those from a study of
several spiral galaxies by Goudfrooij et al.\ (2003). Plots of number
and specific frequency of GCs versus galaxy Hubble type, absolute
magnitude, and stellar mass for our sample show no strong trends and a
modest amount of scatter in GC system properties for otherwise similar
host galaxies.
%for galaxies with similar properties.
%similar host galaxies.

\item We derive the number of blue, metal-poor GCs normalized by the
  galaxy stellar mass ($T_{\rm blue}$) for both NGC~891 and NGC~4013 and
  find them generally 
%
%use the broadband color distributions of the GC candidates
% in each galaxy 
%to estimate the fraction of blue, metal-poor GCs and the
%  galaxy-mass-normalized number of blue GCs ($T_{\rm blue}$) in each
%  galaxy's GC system.  The $T_{\rm blue}$ values for NGC~891 and
%  NGC~4013 are generally 
consistent with a trend of increasing specific frequency of blue GCs
  with increasing galaxy mass.  As was discussed in previous survey
  papers 
%papers from the survey
(e.g., R05, R07), the $T_{\rm blue}$ trend may be a consequence of
  biased, hierarchical galaxy formation.  In such a scenario, a larger
  fraction of the mass of high-mass galaxies is in place when the
  first generation of GCs is formed; the consequence is that
  higher-mass giant galaxies end up with larger $T_{\rm blue}$ values
  than their lower-mass counterparts.
\end{enumerate}

Finally, a key conclusion of this work is that, although this is not
  always possible,
%given the current
%  facilities, 
the most effective way
%  (given current facilities) 
to measure the global properties of the entire system of GCs around
  giant galaxies in the local universe is to combine good-quality
  wide-field imaging 
%in multiple filters 
with $HST$ observations.  The ground-based images allow one to trace a
  GC system to its outer radial extent, while the $HST$ data help to
  characterize contamination and probe further into the galaxy disk.
% toward crowded regions in the galaxy center.  
One can then use both data sets to carefully construct a list of GC
  candidates and a radial profile for the system and derive more
  accurate global quantities that describe the GC population.

\acknowledgments The research described in this paper was supported in
part by an NSF Astronomy and Astrophysics Postdoctoral Fellowship
(award AST-0302095) and by an NSF Faculty Early Career Development
(CAREER) award (AST-0847109) to KLR. We are grateful to the staff at
the WIYN Observatory and Kitt Peak National Observatory for their
assistance during our observing runs. We also thank the anonymous
referee for comments and suggestions that improved the paper.  This
research has made use of the NASA/IPAC Extragalactic Database (NED)
which is operated by the Jet Propulsion Laboratory, California
Institute of Technology, under contract with the National Aeronautics
and Space Administration.

%\appendix
%\section{Floating material and so forth}

%\clearpage

% Now comes the reference list.  In this document, we used \cite to call
% out citations, so we must use \bibitem in the reference list, which
% means we use the LaTeX thebibliography environment.  Please note that
% \begin{thebibliography} is followed by a null argument.  If you forget
% this, mayhem ensues, and LaTeX will say "Perhaps a missing item?" when
% you run it.  Do not call us, do not send mail when this happens.  Put
% the silly {} after the \begin{thebibliography}.
%
% Each reference has a \bibitem command to define the citation format
% to be placed in the text (in []) and the symbolic tag used for
% cross referencing (in {}).
%
% See sample1.tex, or the AASTeX guide, for an alternative to the \cite-
% \bibitem command.

%\clearpage

%\clearpage

%I decided not to label kpc on these figures because for both galaxies
%  there is a question about the distance and kpc would be
%  distance-specific; I think I want the plots to be more general

\clearpage

\tabletypesize{\normalsize}
\begin{deluxetable}{lccrrrr}
\tablecaption{Basic Properties of the Target Galaxies}
\tablewidth{410pt}
\tablehead{\colhead{Name}&\colhead{Type}&\colhead{Inclination}&
  \colhead{$v_{Helio}$} &\colhead{$m-M$} &\colhead{Distance} &\colhead{$M_V^T$}\\
\colhead{} & \colhead{} & \colhead{(deg)} & \colhead{(km s$^{-1}$)} & \colhead{} &
\colhead{(Mpc)} & \colhead{}}
\startdata
\\
NGC~891 & Sb & 84 & 831$\pm$1 & 29.61 & 8.36 & $-$20.79\\ 
NGC~4013 & Sb & 90 & 528$\pm$4 & 30.90 & 15.1 & $-$20.38
\\
\enddata
\tablecomments{Morphological types are from the Hubble Atlas of
  Galaxies (Sandage 1961) for NGC~891 and from RC1 (de~Vaucouleurs \&
  de~Vaucouleurs 1964) for NGC~4013.  Inclination and heliocentric
  radial velocity are from Verheijen \& Sancisi (2001) for NGC~4013.
  Heliocentric radial velocity for NGC~891 is from RC3 (de~Vaucouleurs
  et al.\ 1991) and inclination is from Tully (1988).  Distance to
  NGC~891 is from Tonry et al.\ (2001) (surface brightness
  fluctuations).  Distance to NGC~4013 is from combining the recession
  velocity with respect to the 3~K cosmic microwave background from
  the NASA Extragalactic Database with $H_0$ $=$ 70
  km~s$^{-1}$~Mpc$^{-1}$.  Total absolute magnitudes are from
  combining $V_T^0$ from RC3 with $m-M$.}
\protect\label{table:properties}
\end{deluxetable}

\begin{deluxetable}{llrrr}
\tablecaption{WIYN Minimosaic Observations of the Target Galaxies}
%\tablewidth{400pt}
\tablewidth{300pt}
\tablehead{\colhead{Galaxy} &\colhead{Date}&\multicolumn{3}{c}{Exposure Times (s)} \\
\colhead{} & \colhead{} & \colhead{$B$} & \colhead{$V$} & \colhead{$R$}}
\startdata
NGC~891 & 2001 Jan & 4 x 2100 & 3 x 2000 & 3 x 1500 \\
NGC~4013 & 2001 Jan & 3 x 2100 & 1 x 2000 & 3 x 1500 \\
 & 2009 Mar &  & 4 x 2000 &  
\enddata
\protect\label{table:observations}
\end{deluxetable}

\begin{deluxetable}{lrcl}
\tablecaption{Corrected Radial Profile of the GC System of NGC~891}
%\tablewidth{180pt}
\tablewidth{340pt}
\tablehead{\colhead{Radius} & \colhead{Surface Density} & \colhead{Fractional Coverage} & \colhead{Source}\\
\colhead{(arcmin)} & \colhead{(arcmin$^{-2}$)}}
\startdata
%r(')       N    sfc_dens      serr   goodfrac
0.48 &  9.62 $\pm$ 3.21 & 0.66 & HST\\
0.96 &  4.63 $\pm$  1.89 & 0.43 & HST \\
1.25 &  0.64 $\pm$  0.57 & 0.71 & WIYN\\
1.46 &  0.61 $\pm$  0.61 & 0.36 & HST \\
1.97 &  0.43 $\pm$  0.43 & 0.38 & HST \\
2.17 &  0.91 $\pm$  0.45 & 0.77 & WIYN\\
2.46 &  0.00  $\pm$ 0.00 & 0.42 & HST \\
2.95 &  0.52  $\pm$ 0.37 & 0.41 & HST \\
3.16 &  0.43  $\pm$ 0.33 & 0.64 & WIYN\\
3.44 &  0.37  $\pm$ 0.37 & 0.25 & HST \\
3.94 &  0.00  $\pm$ 0.00 & 0.16 & HST \\
4.15 &  0.10  $\pm$ 0.28 & 0.47 & WIYN\\
4.45 &  0.57  $\pm$ 0.57 & 0.13 & HST \\
5.16 &  0.25  $\pm$ 0.29 & 0.43 & WIYN\\
6.12 &  $-$0.01  $\pm$ 0.25 & 0.32 & WIYN\\
7.12 &  0.09  $\pm$ 0.30 & 0.22 & WIYN\\
8.04 &  $-$0.09 $\pm$ 0.40 & 0.07 & WIYN\\
\tablecomments{For completeness, we list
  all points in the radial profile made using the WIYN + HST GC
  candidate samples; when no GC candidates appear within a particular
  annulus,
%in the radial distribution, 
the surface density is written as ``0.00$\pm$0.00''. 
Negative surface densities sometimes occur because a contamination
correction has been applied to each annulus.}  
\enddata
\label{table:profile n891}
\end{deluxetable}

\begin{deluxetable}{lrcl}
\tablecaption{Corrected Radial Profile of the GC System of NGC~4013}
%\tablewidth{160pt}
\tablewidth{340pt}
\tablehead{\colhead{Radius} & \colhead{Surface Density} & \colhead{Fractional Coverage} & \colhead{Source}\\
\colhead{(arcmin)} & \colhead{(arcmin$^{-2}$)}}
\startdata
%r(')       N    sfc_dens      serr   goodfrac
0.39 & 37.30 $\pm$ 6.93 & 1.00 & HST \\
0.82 &  9.90 $\pm$ 2.65 & 0.72 & HST \\
0.90 &  9.70 $\pm$ 2.31 & 0.72 & WIYN \\
1.34 &  7.07 $\pm$ 2.67 & 0.32 & HST \\
1.81 &  2.22 $\pm$ 1.08 & 0.89 & WIYN \\
2.78 &  0.90 $\pm$ 0.93 & 0.96 & WIYN \\
3.77 &  0.37 $\pm$ 0.89 & 0.96 & WIYN\\
4.68 &  0.36 $\pm$ 0.90 & 0.66 & WIYN\\
\enddata
\label{table:profile n4013}
\end{deluxetable}

\begin{deluxetable}{lcccc}
\tablecaption{Coefficients from Fitting Radial Profile Data}
\tablewidth{370pt}
\tablehead{
\colhead{} & \multicolumn{2}{c}{de~Vaucouleurs Law} &
\multicolumn{2}{c}{Power Law}\\
\cline{2-3} \cline{4-5}\\
\colhead{Galaxy} & \colhead{a0}& \colhead{a1} & \colhead{a0}& \colhead{a1}}
\startdata
NGC~891 & 3.02 $\pm$ 0.33 & $-$2.53 $\pm$ 0.29 & 0.45 $\pm$ 0.07 & $-$1.62 $\pm$ 0.18\\
NGC~4013 & 4.02 $\pm$ 0.30 & $-$3.11 $\pm$ 0.32 & 0.88 $\pm$ 0.05 & $-$1.73 $\pm$ 0.18\\
\enddata
\protect\label{table:coefficients}
\end{deluxetable}

\begin{deluxetable}{lrccc}
\tablecaption{Total Numbers and Specific Frequencies of the GC Systems}
\tablewidth{290pt}
\tablehead{\colhead{Galaxy}
&\colhead{$N_{GC}$}&\colhead{$S_N$} 
&\colhead{$T$}&\colhead{$T_{\rm blue}$}\\
\colhead{} & \colhead{} & \colhead{} & \colhead{} & \colhead{}}
\startdata
NGC~891 & 70$\pm$20 & 0.3$\pm$0.1 & 0.6$\pm$0.3  & 0.3$\pm$0.2 \\
NGC~4013 & 140$\pm$20 & 1.0$\pm$0.2 & 1.9$\pm$0.5 & 1.4$\pm$0.3 \\
\enddata
\protect\label{table:total numbers}
\end{deluxetable}

\end{document}